\documentclass[letterpaper,twocolumn,10pt]{article}

\usepackage[utf8]{inputenc}
\usepackage[english]{babel}
\usepackage{csquotes}
\PassOptionsToPackage{table,xcdraw}{xcolor}
\usepackage{makecell}

\usepackage[style=numeric-comp, maxcitenames=1, mincitenames=1, maxbibnames=3, minbibnames=3, natbib=true, backend=biber]{biblatex}
\AtBeginBibliography{\small}

\usepackage{usenix-2020-09}

\usepackage{tikz}
\usepackage{amsmath, stmaryrd}
\usepackage{nopageno}

\usepackage{bbold}
\usepackage{algorithm}
\usepackage[noend]{algpseudocode}
\usepackage{multirow}
\usepackage{authblk}
\definecolor{mygray}{gray}{0.5}

\newcommand\CommentLine[1]{{\color{mygray} \textit{#1}}}
\newcommand{\PUB}{_\text{pub}}
\newcommand{\BASE}{_\text{base}}
\newcommand{\ATT}{_\text{att}}
\newcommand{\argmax}{\operatornamewithlimits{argmax}}
\newcommand{\given}{\ | \ }
\newcommand*\diff{\mathop{}\!\mathrm{d}}

\begin{document}

\date{}

\title{\Large \bf Reconstructing training data from document understanding models}

\author[1, 2]{Jérémie Dentan}
\author[1]{Arnaud Paran}
\author[1]{Aymen Shabou}

\affil[1]{Crédit Agricole SA}
\affil[2]{École Polytechnique, IP Paris}

\maketitle
 
\begin{abstract}
Document understanding models are increasingly employed by companies to supplant humans in processing sensitive documents, such as invoices, tax notices, or even ID cards. However, the robustness of such models to privacy attacks remains vastly unexplored.

This paper presents CDMI, the first reconstruction attack designed to extract sensitive fields from the training data of these models. We attack LayoutLM and BROS architectures, demonstrating that an adversary can perfectly reconstruct up to 4.1\% of the fields of the documents used for fine-tuning, including some names, dates, and invoice amounts up to six-digit numbers. When our reconstruction attack is combined with a membership inference attack, our attack accuracy escalates to 22.5\%.

In addition, we introduce two new end-to-end metrics and evaluate our approach under various conditions: unimodal or bimodal data, LayoutLM or BROS backbones, four fine-tuning tasks, and two public datasets (FUNSD and SROIE). We also investigate the interplay between overfitting, predictive performance, and susceptibility to our attack. We conclude with a discussion on possible defenses against our attack and potential future research directions to construct robust document understanding models.

\end{abstract}


\begin{figure}[t]
    \centering
    \includegraphics[width=\columnwidth]{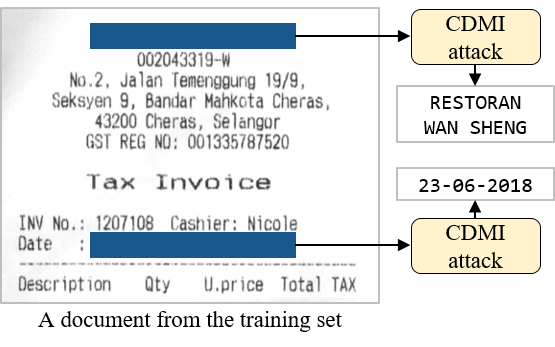}
    \caption{A document (licensed CC BY 4.0 DEED by \citet{huang_icdar2019_2019}) where two fields are perfectly reconstructed by CDMI. A model with LayoutLM architecture \cite{xu_layoutlm_2019} is trained on SROIE dataset \cite{huang_icdar2019_2019}. Then, when the date or the company is scrubbed, the adversary is able to reconstruct it.}
    \label{fig:teaser}
\end{figure}

\section{Introduction}

Document understanding models aim at processing visually rich documents, such as handwritten forms, tax invoices or scanned tables, where information is encoded both in textual content and layout. Consequently, most document understanding models adapt a language model architecture to make it layout-aware \cite{xu_layoutlm_2019, xu_layoutlmv2_2021, huang_layoutlmv3_2022, hong_bros_2021, llados_lambert_2021, powalski_going_2021}. These models have countless real-world applications and are  employed by numerous companies for tasks extending from key information extraction \cite{luo_geolayoutlm_2023, huang_layoutlmv3_2022, park_cord_2019} to document classification \cite{nikolentzos_message_2020, harley_evaluation_2015}, and question answering \cite{wang_layout_2023, mathew_docvqa_2021}. 

Moreover, a range of neural network architectures have been found vulnerable to membership inference or reconstruction attacks in multiple domains: computer vision models \cite{fredrikson_privacy_2014, fredrikson_model_2015, zhang_secret_2020, khosravy_model_2022, kahla_label-only_2022, wang_variational_2022}, language models \cite{zhang_text_2022, carlini_extracting_2021, carlini_quantifying_2022, lee_language_2022, shao_quantifying_2023, lukas_analyzing_2023, mireshghallah_quantifying_2022, lehman_does_2021}, graph models \cite{zhang_inference_2022, olatunji_membership_2021}, and diffusion models \cite{carlini_extracting_2023}, among the most popular. Those attacks enable an adversary to obtain information on the training data, posing a serious threat to confidentiality (see section \ref{sec:background_privacy_attacks} and \cite{shokri_membership_2017, salem_let_2023}).

Surprisingly, despite their resemblance to language \mbox{models} and vision models, we have not discovered any existing studies on the robustness of document understanding models to such privacy attacks (at the time of writing, in October 2023). The most related research paper we found in this domain is by \citet{meehan_sentence-level_2022}. However, they define a document as a collection of sentences, excluding any layout information. Thus, the model they target (Sentence-BERT \cite{reimers_sentence-bert_2019}) differs significantly from state-of-the-art document understanding models. The reconstruction attack we present in this paper is the first to target some of the most commonly used layout-aware models and treat documents as multimodal data.

\subsection{Contributions}

Our main contributions are as follows:

\begin{itemize}
    \item We develop the first reconstruction attack that targets layout-aware document understanding models: a \textit{white-box} attack named CDMI (Combinatorial Document Model Inversion), designed to target unimodal or bimodal models. We employ CDMI to reconstruct fields from the fine-tuning datasets of our models, although it is also readily adaptable for pre-training datasets.
    
    \item We combine CDMI with existing membership inference attacks, resulting in an end-to-end attack where the adversary extracts arbitrary fields in the training dataset.
    
    \item We introduce two new end-to-end metrics to evaluate simultaneously the reconstruction and the membership inference attack. Our intention is to evaluate the full extent of potential harm that an adversary could cause.

    \item We successfully evaluated our attack in various settings, making it one of the few attacks capable of reconstructing non-synthetic data from an encoder-only model. These settings include unimodal or bimodal models, with either LayoutLM v1 \cite{xu_layoutlm_2019} or BROS \cite{hong_bros_2021} backbone, using four different training tasks, and two public fine-tuning datasets: FUNSD \cite{jaume_funsd_2019} and SROIE~\cite{huang_icdar2019_2019}. 

    \item We demonstrate that the success of our attack is not due to over-fitting or data duplication, and that memorization occurs early in the training pipeline. We also prove that both the layout and the visual modality contribute to memorization, proving that documents should be considered as a specific data type with dedicated attacks.
\end{itemize}

\section{Background and related work} \label{sec:background}

\subsection{Document understanding models} \label{sec:background_doc_models}

A document processing pipeline aims at extracting meaningful information from raw document images such as scans, pictures, or screenshots. Deep learning-based pipelines usually include two main steps \cite{subramani_survey_2021}: 

\begin{itemize}
    \item An Optical Character Recognition (OCR) is used to extract the textual content of the document: its output typically includes \textit{words} and their associated \textit{bounding boxes}, which denote the coordinates of the four points encompassing the word. Then, the words are fed to a \textit{tokenizer}. It splits the words into \textit{tokens}, which are smaller pieces of text that are part of a specific \textit{vocabulary}.
    \item These tokens and their bounding boxes are processed by a \textit{document understanding model}, which extracts specific information from them. A model is said to be \textit{layout-aware} when it uses both the tokens and the layout information derived from the bounding boxes \cite{wang_layout_2023}. 
\end{itemize}

\paragraph{The architectures we target: LayoutLM v1 and BROS}

Recently, OCR-free pipelines were developed \cite{dhouib_docparser_2023, guo_eaten_2019, cheng_trie_2022, kim_ocr-free_2022}. However, our focus will be on deep learning models specifically designed to work with an OCR, given that OCR-free approaches are relatively new and rarely deployed in production. The category of OCR-based models includes recent architectures with state-of-the-art results such as LayoutLM v1 \cite{xu_layoutlm_2019}, LayoutLM v2 \cite{xu_layoutlmv2_2021}, LayoutLM v3 \cite{huang_layoutlmv3_2022}, BROS \cite{hong_bros_2021}, LAMBERT \cite{llados_lambert_2021}, TILT \cite{powalski_going_2021}, and DocFormer \cite{appalaraju_docformer_2021}.

We decided to attack two different architectures: LayoutLM v1 and BROS, for multiple reasons. Firstly, both architectures demonstrate strong performance and are frequently used in real-world applications.\footnote{As of October 2023, LayoutLM v1 had  over 29M downloads: \url{https://huggingface.co/api/models/microsoft/layoutlm-base-uncased?expand[]=downloadsAllTime}} In addition, their pretrained weights are available under permissive licenses (in contrast to LayoutLM v3), making them particularly suitable for commercial applications. Thus, "LayoutLM" will implicitly refer to LayoutLM v1 in the following.

\paragraph{Transformer-based document encoders} These architectures are transformer-based encoder-only models, akin to BERT \cite{devlin_bert_2019}. They employ multiple layers of multi-headed Transformers \cite{vaswani_attention_2017} to embed each input token in a feature space of dimension $d = 768$. However, unlike BERT, these architectures use a 2-dimensional spatial encoding to consider the layout information provided by the bounding boxes.

In addition to the tokens and their bounding boxes, a document understanding model can exploit the raw images of the documents \cite{xu_layoutlm_2019, powalski_going_2021, hong_bros_2021}. In the case of LayoutLM and BROS, visual features are generated by a computer vision model, and subsequently added to the textual embedding of each token using Region Of Interest alignment \cite{he_mask_2017}. We will refer to these models as \textit{bimodal}, in contrast to the simpler layout-aware models which we will refer to as \textit{unimodal}. 

\paragraph{Training objectives} LayoutLM and BROS encoders are often used in a transfer learning setting. First, the encoders are \textit{pre-trained} on a very large corpus of documents, the IIT-CDPI Test Collection 1.0 \cite{lewis_building_2006}. The main pre-training task is Masked Language Modeling (MLM), where approximately 15\% of the tokens are masked, and the model is trained to reconstruct them. Afterwards, the model is \textit{fine-tuned} on a specific task by keeping the pretrained backbone and replacing the last layers with a new classification head specifically designed for this task. Our attack will target these fine-tuned models.

We evaluated our attack against models fine-tuned on three common Key Information Extraction (KIE) tasks, which were previously employed by \citet{hong_bros_2021} to evaluate BROS:

\begin{itemize}
    \item Entity Extraction with BIO tagging (EE-BIO): it aims at extracting some fields of the document by classifying its tokens as "Beginning", "Inside" or "Out" of an entity~\cite{alshammari_impact_2021}.
    \item Entity Extraction with SPADE tagging (EE-SPADE or EE-SPD): it also aims at extracting some fields using the SPADE tagging \cite{hwang_spatial_2021}. In a nutshell, the model is trained to identify the predecessor of each token in the entities of the document.
    \item Entity Linking (EL): it aims at establishing connections between the entities using their semantic relations. We also used SPADE tagging for this task \cite{hwang_spatial_2021}.
\end{itemize}

In addition to those three tasks, we also attacked models that were fine-tuned on the same MLM task that is utilized during pre-training. This scenario does not correspond to a real-world setting; however we did this to compare the susceptibility of the KIE and the MLM tasks under the same conditions (transfer learning with the same fine-tuning dataset, etc.). All these tasks ultimately involve a token-level classification task, for which we used a cross-entropy loss.

\subsection{Privacy attacks} \label{sec:background_privacy_attacks}

In this section, we introduce privacy attacks and review the existing ones in the field of Natural Language Processing (NLP). Indeed, the architectures of language models are similar to those of document understanding models, for which such attacks have not yet been studied.

\paragraph{Taxonomy} Various types of attacks target the confidentiality of models trained on sensitive data by exploiting  unintended memorization of their training set. Following \cite{lukas_analyzing_2023}, we distinguish three levels of privacy attacks:

\begin{itemize}
    \item \textit{Extraction attacks}: the adversary has only access to the model, and seeks to extract as much data as feasible from the training set. In NLP, such attacks are studied in~\cite{carlini_extracting_2021, carlini_quantifying_2022, lee_language_2022, shao_quantifying_2023, lukas_analyzing_2023, zhang_text_2022, yu_bag_2023}.
    \item \textit{Reconstruction attacks}: the adversary has access to a model and a context such as the prefix of a sentence, or a document with a scrubbed field, and seeks to reconstruct the missing parts. In NLP, attacks of this kind include~\cite{lukas_analyzing_2023, parikh_canary_2022, elmahdy_privacy_2022, meehan_sentence-level_2022}.
    \item \textit{Membership inference attacks} \cite{shokri_membership_2017}: the adversary has access to data samples, and attempts to predict whether they are part of the model's training set or not. Formally, this is equivalent to a reconstruction attack with a finite list of candidates for the reconstruction. Notable examples of these attacks are \cite{mireshghallah_quantifying_2022, lehman_does_2021, song_information_2020, mahloujifar_membership_2021, rezaei_accuracy-privacy_2023}.
\end{itemize}

In this paper, we focus on reconstruction attacks. We begin with a ground truth document, represented for now as a sequence of tokens $d_g$ in a vocabulary $\mathcal{V}$. A field is scrubbed by replacing it with the special token \texttt{[MASK]} to form $\widetilde{d_g}$. Then, the adversary has access to $\widetilde{d_g}$ and optimizes equation \ref{eq:recon_optim} to reconstruct the scrubbed fields.

\begin{equation} \label{eq:recon_optim}
    \argmax_{d \in \mathcal{V}^*}\ P_\theta(d \given \widetilde{d_g})
\end{equation}

Here, $P_\theta$ denotes the likelihood of $d$ being a good reconstruction of $\widetilde{d_g}$ given the model's weights $\theta$. However, this probability is frequently either ill-defined or intractable. Hence, practical attacks necessitate the construction of a suitable proxy for $P_\theta$  (see \cite{lukas_analyzing_2023} and sections \ref{sec:autoregressive_formulation} and \ref{sec:token_level_optim}).

\paragraph{Attacks against decoder-only models} In the context of extraction and reconstruction attacks in NLP, a significant difference exists between decoder-only models and others \cite{ishihara_training_2023, yang_harnessing_2023}. Decoder-only models, such as the ones of the GPT family, are trained in an auto-regressive way to generate the next token in a sequence given its prefix. They are well-designed to generate fluent content. 

Existing attacks against such models directly use this generation ability to extract training data \cite{carlini_extracting_2021, carlini_quantifying_2022, lee_language_2022, lukas_analyzing_2023, yu_bag_2023}. These attacks reconstruct data from left to right. At step $t$, using the previously reconstructed prefix sequence $\widetilde{s_g} = c_1 \dots c_{t-1}$, they attempt to reconstruct token $c_t$ . As a result, if they directly replace $P_\theta$ by the probability distribution calculated by the model for the next token, equation \ref{eq:recon_optim} ends up being exactly the one the model is trained to solve:

\begin{equation} \label{eq:recon_optim_decoded}
    \argmax_{c_t\in \mathcal{V}}\ P_\theta(c_t \given c_1, \dots, c_{t-1})
\end{equation}

\paragraph{Attacks against encoder-only model} Encoder-only models, such as LayoutLM or BROS, are not designed to compute a probability distribution over sentences. This makes it difficult for them to generate fluent content \cite{wang_bert_2019, goyal_exposing_2022}. This is a crucial difference, because this ability is needed to reconstruct plausible content in equation \ref{eq:recon_optim}. However, various strategies have been proposed to attack encoder-only models in NLP:

\begin{itemize}
        \item \citet{zhang_text_2022} employ an auxiliary GPT-2 model \cite{budzianowski_hello_2019} to generate content. They utilize the PPLM mechanism \cite{dathathri_plug_2019} to increase the likelihood of generating content from the training set of the BERT model they target~\cite{devlin_bert_2019}.
        \item \citet{lehman_does_2021} focus on small prompts in a clinical context, and conclude that their attack does not significantly expose the training data. \citet{mireshghallah_quantifying_2022} enhance their attack in a membership inference setting, adopting the energy-based interpretation of \cite{goyal_exposing_2022} instead of Gibbs sampling to compute text fragment likelihoods.
        \item \citet{carlini_quantifying_2022} assess their attack against T5 model \cite{raffel_exploring_2020}. However, they addressed a simpler scenario where it is not needed to generate fluent content: 15\% of randomly chosen tokens are masked and need to be reconstructed.
        \item \citet{song_information_2020} and \citet{parikh_canary_2022} formulate a continuous relaxation of equation \ref{eq:recon_optim} inspired from the work of \citet{jang_categorical_2017} for its resolution.
        \item \citet{mahloujifar_membership_2021} target embedding models, exploiting specific properties of these models in their attack.
        \item \citet{elmahdy_privacy_2022} focus on canary reconstruction and use an autoregressive proxy. They decode tokens from left to right, with the objective of minimizing the loss of the model. \citet{elmahdy_deconstructing_2023} improves the approach of \cite{elmahdy_privacy_2022} by using a MLM head to select candidates before evaluating the loss of the target model.
        \item \citet{meehan_sentence-level_2022} only focus on differential privacy \cite{dwork_calibrating_2006, dwork_algorithmic_2014, dwork_firm_2011, abadi_deep_2016}, demonstrating theoretical privacy bounds without implementing the attack to test these limits.
\end{itemize}

Thus, reconstructing training data from encoder-only models is a difficult task, and there is no standard method for it. To attack real-world document understanding models, we could not directly apply existing methods. Indeed, they either constituted membership inference attacks (which are less complex than reconstructions) \cite{lehman_does_2021, mireshghallah_quantifying_2022}, were only evaluated on frequently repeated canaries \cite{elmahdy_deconstructing_2023, elmahdy_privacy_2022, parikh_canary_2022}, or were inapplicable to our scenario \cite{mahloujifar_membership_2021, carlini_quantifying_2022, meehan_sentence-level_2022, song_information_2020}.

We selected and combined promising ideas from various existing attack, improving and optimizing them for our scenario. We utilized an autoregressive proxy as in \cite{elmahdy_privacy_2022, elmahdy_deconstructing_2023}, and leveraged an auxiliary generator to enhance the fluency of the reconstructions as in \cite{zhang_text_2022, elmahdy_deconstructing_2023}. We also developed a customized method for incorporating the loss of the target model, which we combined with existing membership inference metrics \cite{carlini_extracting_2021}. Finally, we integrated the visual modality in our attack to extract information memorized by the visual encoder. This led to our hybrid attack, CDMI, the first of its kind capable of reconstructing real documents.

\paragraph{Defense against privacy attacks} According to \cite{ishihara_training_2023}, the main defense techniques in a centralized setting are:

\begin{itemize}
    \item \textit{Data sanitization} \cite{ren_recon_2016, continella_obfuscation-resilient_2017, vakili_downstream_2022, brown_what_2022}. It involves removing personal data from the training set. However, this technique is limited due to the context-dependent definition of personal data \cite{brown_what_2022}. Deduplication, which consists in eliminating duplicate data, can also be useful as duplicated data is more likely to be memorized \cite{kandpal_deduplicating_2022}.
    \item \textit{Differential Privacy (DP)} \cite{dwork_calibrating_2006, dwork_firm_2011, dwork_algorithmic_2014, abadi_deep_2016}. This training paradigm adds noise to each gradient during the training phase up to a certain level, to safeguard every sample in the training set. However, the effectiveness of DP in genuine real-world settings is disputed \cite{domingo-ferrer_limits_2021, tramer_considerations_2022}.
    \item \textit{Regularization.} Although not necessary, overfitting facilitates the memorization of training data \cite{yeom_privacy_2018, zhang_understanding_2021, feldman_does_2020, carlini_extracting_2021}. Hence, some mechanisms involving regularization have been proposed \cite{mireshghallah_privacy_2021, hu_membership_2022}.
    \item \textit{Post-processing.} Confidence masking and output filtering can be employed during post-processing \cite{fredrikson_model_2015, perez_red_2022}.
\end{itemize}

\section{Threat model} \label{sec:threat_model}

In this section we introduce our \textit{threat model}, which defines the assumptions we made for the development of our attack.

\subsection{Adversarial capabilities and goals} \label{sec:adv_goals}

We make two primary assumptions about the adversary's capabilities:

\begin{itemize}
    \item \textit{White box hypothesis.} The adversary is supposed to have white-box access to the model, meaning complete access to its architecture and weights. Concretely, our attack is gradient-free, but because it requires computing the token-level loss for a large number of inputs, it is impractical in standard black-box environments. This is why we classify it as a white-box attack.
    \item \textit{Scrubbed data.} Given that we are developing a reconstruction attack, we assume the adversary has access to scrubbed data. In our context, this means that the adversary can access a document (token, bounding boxes, and the raw image) where a certain field has been masked (tokens replaced by \texttt{[MASK]}, and a white patch to replace the field in the image). This implies that the adversary knows the number of tokens in the target field. While this is a strong supposition, it is necessary to make our optimization in equation \ref{eq:recon_optim} tractable (see III.B in \cite{lukas_analyzing_2023}). Moreover, this assumption is fairly realistic for documents in which the fields adhere to strict rules (e.g. IBANs, credit card number and expiration date, etc.).
\end{itemize}

We focus on an adversary whose goal is to reconstruct the textual modality of documents, regardless of whether the target model is unimodal or bimodal. Indeed, for real-world document understanding models, sensitive information is often encoded in the textual modality. For example, for identity theft, reconstructing the ID card number and expiration date is significantly more valuable than the general layout of the card, which is common knowledge.

More specifically, we distinguish two variants of our attack, a \textit{one-shot} one and a \textit{multi-shot} one, each with slightly different adversarial goals:

\begin{itemize}
    \item \textit{One-shot variant.} In this scenario, the adversary makes one reconstruction attempt per field in the datasets. Here, their objective is to maximise the similarity metrics between their reconstruction and the ground truth.
    \item \textit{Multi-shot variant.} This setting is closer to a real-world scenario. The adversary makes multiple attempts against each field, and ultimately uses a membership inference metric to filter the most plausible reconstructions, as referenced in \cite{carlini_extracting_2021, yu_bag_2023}). The adversary has two objectives. First, to generate high-quality reconstructions. Second, to accurately retain the correct ones using the membership inference metric (see section \ref{sec:eval_multishot}).
\end{itemize}

\subsection{The reconstruction game} \label{sec:reconstruction_game}

\begin{table}[b!]
\begin{tabular}{m{0.2\columnwidth} m{0.67\columnwidth}}
\hline
Notation                          & Description                                        \\ \hline
$\mathcal{T}$                     & A stochastic training procedure                    \\
$\mathcal{A}$                     & A procedure denoting an adversary                  \\
$\mathcal{V}$                     & The vocabulary for the tokens                      \\
$\mathcal{I} = \mathbb{R}^{3 \times d_i}$  & The RGB space of documents' images      \\
$D \sim \mathcal{D}^n$            & Sample $n$ docs from a distribution $\mathcal{D}$ on $\mathcal{V}^* \times \mathcal{I}$\\
$d \in \mathcal{V}^* \times \mathcal{I}$ & A document (sequence of tok. + image)       \\
$\widetilde{d} \in \mathcal{V}^* \times \mathcal{I}$ & A scrubbed document derived from $d$               \\
$y \in \mathbb{R}^*$              & A classification label for a document              \\
$f \in \mathcal{V}^*$             & A field of $d$ containing personal info.           \\
$\widetilde{f} \in \mathcal{V}^*$ & A reconstruction attempt for $f$                   \\
$k = |f|$                         & The number of tokens in $f$                         \\ \hline
\end{tabular}
\caption{Notations}
\label{tab:game_notations}
\end{table}

Following \cite{salem_let_2023, wu_methodology_2016}, we define the \textit{reconstruction game}, that serves to formalize the threat model, the information the adversary can access, and what will be used for evaluation. It is an adaptation of the one in \citet{lukas_analyzing_2023}. In addition to the notations of table \ref{tab:game_notations}, we define:

\begin{itemize}
    \item $\text{EXTRACT}(d)$: extracts the list of fields of $d$ that contain personal information.
    \item $\text{SCRUB}(d, f) \rightarrow \widetilde{d}, y$: scrubs a field $f$ in a document $d$ by replacing it by \texttt{[MASK]}, and returns its classification label $y$ (which is public information).
\end{itemize}

Our multi-shot reconstruction game is presented in algorithm \ref{alg:game_multishot}. Initially, a model is trained on a private dataset $D$ (lines 2--3). Then, for each document in the dataset (line 6), and for each field in this document (line 7), the field is scrubbed and the adversary attempts $n_a$ reconstructions (lines 8--12). The adversary has access to the ground truth number of tokens $k$ (see section \ref{sec:adv_goals}). They also know $\mathcal{D}$, $\mathcal{T}$, and $n$, which means they can train auxiliary models on the same distribution. Then, a Membership Inference (MI) metric is utilized to keep the best reconstruction attempt for each field (line 13). Subsequently, the MI metric is used to sort the fields by probability of successful reconstruction (line 16). Finally, we evaluate the attempts, examining both their similarity to their ground truth, and their order with respect to the MI metric computed at line 16 (see section \ref{sec:eval_multishot}).

The one-shot reconstruction game is very similar. Since there is only one attempt per scrubbed field, lines 9, 10, 12, 13 are removed. Furthermore, the order of the fields is not considered during evaluation, so line 16 is removed.

With these notations, we can refine equation \ref{eq:recon_optim} introduced earlier. At line 11, the adversary seeks to solve equation \ref{eq:detailed_recon_optim}. We recall that the adversary only tries to reconstruct the textual modality of the fields, which is why the optimization is over $\widetilde{f} \in \mathcal{V}^k$ and not over $\widetilde{f} \in \mathcal{V}^k \times \mathcal{I}$ (see section \ref{sec:adv_goals} and the paragraph on visual modality in section \ref{sec:autoregressive_formulation}).

\begin{equation} \label{eq:detailed_recon_optim}
    \mathcal{A}(\mathcal{T}, \mathcal{D}, n, \theta, \widetilde{d}, y, k) = \argmax_{\widetilde{f} \in \mathcal{V}^{k}}\ P_\theta(\widetilde{f} \given \widetilde{d}, y)
\end{equation}

\begin{algorithm}[t!]
\caption{The multi-shot reconstruction game}\label{alg:game_multishot}
\algrenewcommand\algorithmicprocedure{\textbf{experiment}}
\begin{algorithmic}[1]
\Procedure{RECON\_M\_SHOT}{$\mathcal{D}, \mathcal{T}, \mathcal{A}, n$}
    \Statex \CommentLine{Sample docs and train a model}
    \State $D \sim \mathcal{D}^n$
    \State $\theta \gets \mathcal{T}(D)$
    \State $F \gets [\ ]$
    \State $\widetilde{F} \gets [\ ]$
    \Statex \CommentLine{$n_a$ reconstruction attempts on each field}
    \For{$d \in D$}
        \For{$f \in \text{EXTRACT}(d)$}
            \State $\widetilde{d}, y \gets \text{SCRUB}(d, f)$
            \State $\text{attempts} \gets [\ ]$
            \For{$a \in [1, \dots, n_a]$}
                \State $\widetilde{f} \gets \mathcal{A}(\mathcal{T}, \mathcal{D}, n, \theta, \widetilde{d}, y, k)$
                \State $\text{attempts} \gets \text{CONCAT}(\text{attempts}, [\ \widetilde{f}\ ])$
            \EndFor
    \Statex \CommentLine{MI to keep one attempt per field}
            \State $\widetilde{f} \gets \text{MI\_FILTER}(\text{attempts})$
            \State $F \gets \text{CONCAT}(F, [\ f\ ])$
            \State $\widetilde{F} \gets \text{CONCAT}(\widetilde{F}, [\ \widetilde{f}\ ])$ 
        \EndFor
    \EndFor
    \Statex \CommentLine{MI to sort reconstructions by plausibility}
    \State $F, \widetilde{F} \gets \text{MI\_SORT}(F, \widetilde{F})$
    \State $\text{EVAL}(F, \widetilde{F})$
\EndProcedure
\end{algorithmic}
\end{algorithm}

\subsection{Ethical considerations} \label{sec:ethics}

This paper presents an attack that specifically targets the privacy of document understanding models. This raises ethical considerations because such models are sometimes trained on private data and used in real-world settings.

First, we reduce ethical concerns by only attacking models trained on publicly-available data: LayoutLM v1 \cite{xu_layoutlm_2019} and BROS \cite{hong_bros_2021} are both pre-trained on the public dataset IIT-CDIP Test Collection 1.0 \cite{lewis_building_2006}. Moreover, we fine-tune them on two other public datasets, FUNSD \cite{jaume_funsd_2019} and SROIE~\cite{huang_icdar2019_2019}.

Moreover, our attack still necessitates a substantial amount of prior information to successfully reconstruct data. As for every reconstruction attack, this includes access to scrubbed data, an uncommon scenario. Consequently, while our approach may set the foundation for more efficient methods with less information, we argue that the advantages of disclosing it outweigh potential harm. Researchers and companies training document understanding models on private data must understand that robust attacks are very likely to exist in the future. Therefore, it is crucial they protect access to their trained models with the same stringency as their databases.

Finally, we believe that disclosing our attack is important to shed light on privacy attacks and ensure they are considered by institutional regulators. Currently, despite the development of attacks on numerous model types, the fact that trained models may leak personal information is frequently overlooked in data regulation texts such as  RGPD and AI Act in European Union, APPI in Japan, CCPA in California, etc.

\section{The CDMI reconstruction method} \label{sec:cdmi_method}

The section details the methodology employed for our reconstruction attack: CDMI (Combinatorial Document Model Inversion). Specifically, it elaborates on three aspects: the chosen proxy for $P_\theta$ in equation \ref{eq:detailed_recon_optim}; how we optimize it; and the membership inference metric we used.

In summary, our attacks proceeds as follows. First, we approximate the probability distribution over the fields using an autoregressive proxy, meaning that tokens will be reconstructed from left to right within each field (see section \ref{sec:autoregressive_formulation}). To reconstruct each token, we compute and optimize a token-level probability $P_\theta^\text{tok}$ (see section \ref{sec:token_level_optim}). For this, we employ a masked model trained on public data to select candidates. Next, we evaluate the loss of the target model for each candidate, using these computations to approximate $P_\theta^\text{tok}$. Finally, we sample the reconstruction from this approximate distribution. We also present two variants of our attack: the one-shot one, and the multi-shot one, where the adversary attempts several reconstructions against each field and selects the best one using a membership inference metric (see section \ref{sec:variants}).

\subsection{An autoregressive proxy} \label{sec:autoregressive_formulation}

To solve equation \ref{eq:detailed_recon_optim}, the adversary needs to choose a convenient proxy for $P_\theta$, that should correctly represent the likelihood of a document being in the training set while being easy to optimize. Adapting the methodology of \cite{elmahdy_privacy_2022} in NLP,  we employ an \textit{autoregressive formulation}, meaning that scrubbed tokens are sequentially reconstructed from left to right.

Equation \ref{eq:optim_autoreg_proxy} presents this autoregressive proxy, with notations derived from table \ref{tab:game_notations}. We recall that $ k$, the number of tokens in the target field, is known to the adversary (see section \ref{sec:adv_goals}). This formula employs a token-level proxy $P_\theta^\text{tok}$, that represents the probability that the selection of token $v_t \in \mathcal{V}$ at step $t$ results in a good reconstruction. 

\begin{equation} \label{eq:optim_autoreg_proxy}
P_\theta(\widetilde{f} \given \widetilde{d}, y)\ =\  \prod_{t=1}^{k} P_\theta^{\text{tok}}(v_t \given \widetilde{d}, y, v_1, \dots,  v_{t-1})
\end{equation}

Although encoder-only models such as LayoutLM or BROS are not designed for such an autoregressive formulation \cite{wang_bert_2019, goyal_exposing_2022}, we used this approximation for several reasons:

\begin{itemize}
    \item This autoregressive modeling substantially decreases the size of the search space. Rather than $\mathcal{V}^k$, our optimization problem is divided into $k$ sub-problems in $\mathcal{V}$.
    \item More complex methods require a significantly higher number of model calls \cite{goyal_exposing_2022}. Considering that our attack is already computation-intensive, our autoregressive proxy serves to maintain this number to a minimum.
    \item The left-to-right decoding is requisite for the SPADE \cite{hwang_spatial_2021} implementation we used. Indeed, each token is trained to point towards its predecessor, thereby making it significantly easier to decode if the predecessor has already been reconstructed.
    \item Using an autoregressive decoding enables us to use various heuristics derived from reconstruction attacks on decoder-only models \cite{yu_bag_2023}.
\end{itemize}

As explained in section \ref{sec:background_doc_models}, our attack targets both unimodal (text + layout) and bimodal (text + layout + image) models. Nevertheless, the objective of the adversary is solely to reconstruct the textual data (see section \ref{sec:adv_goals}). When attacking a bimodal model, we did the following:

\begin{itemize}
    \item To scrub bimodal data, target tokens are replaced with \texttt{[MASK]} token, and a white patch masks the content of the field in the image.
    \item With our autoregressive proxy, we sequentially replace the target tokens with their reconstruction in the textual modality.
    \item The visual modality remains masked throughout the entire field reconstruction process. This means that the white patch masks all tokens, and remains unchanged even when the initial ones have already been reconstructed in the visual modality. Indeed, the adversary makes no attempt to reconstruct the visual modality (cf. section \ref{sec:adv_goals}), even when attacking a bimodal model.
\end{itemize}

This is why our equations only optimize on $\mathcal{V}^k$. The information of the image $\mathcal{I}$ comes from the scrubbed document $\widetilde{d} \in \mathcal{V}^* \times \mathcal{I}$.

\subsection{A token-level combinatorial optimization} \label{sec:token_level_optim}

\begin{figure}[ht]
    \centering
    \includegraphics[width=0.93\columnwidth]{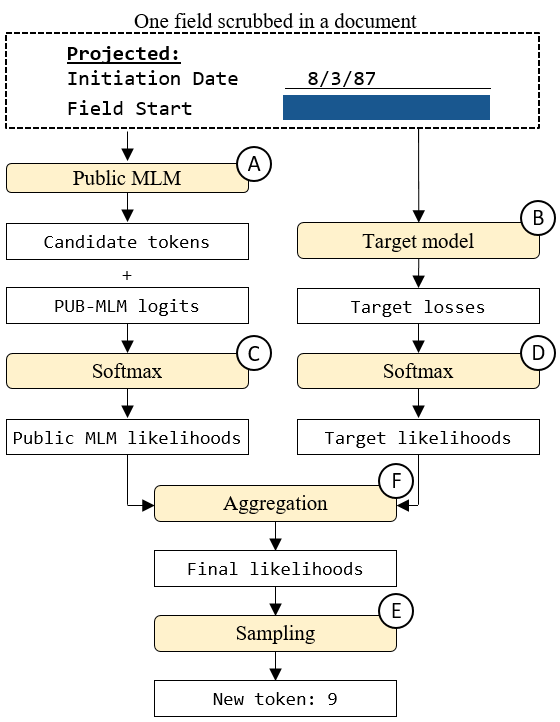}
    \caption{Computing and maximizing $P_\theta^\text{tok}$ to invert a token. We first use a masked model trained on public data to select $N_c = 128$ candidates (A). Then, we compute the loss of the target model with each candidate (B), and aggregate these losses to obtain a probability distribution over the candidates (C, D, E). Finally, we sample the reconstructed token from this distribution (F), and repeat this process for the next tokens.}
    \label{fig:proxy_optimization}
\end{figure}

The computation and optimization of $P_\theta^\text{tok}$ involves several steps that are shown in algorithm \ref{alg:token_level_proxy} and figure \ref{fig:proxy_optimization}. In essence, we leverage an auxiliary MLM head trained on public data to choose a list of potential tokens for reconstruction. Then, we compute the loss of the target model for each candidate, and select the one leading to the minimal loss. This is why we refer to our attack as \textit{combinatorial}. Unlike the continuous relaxation used in NLP by \cite{parikh_canary_2022, song_information_2020}, we use discrete optimization by testing each candidate.

\paragraph{Details on step 1 (lines 2--4 in algo. \ref{alg:token_level_proxy}, block A in fig. \ref{fig:proxy_optimization})} This step employs an auxiliary MLM head trained with $\mathcal{T}$ and $ \mathcal{D}$. These are assumed to be public knowledge, hence this head is referred to as PUB-MLM. The step consists in a single forward-pass of the document in PUB-MLM, followed by the selection of the top-$N_c$ most plausible candidates. 

This first step is needed because we empirically observed that the token minimizing the loss of the target model is often not the one we seek (e.g. unexpected words like "paranoia" for a "date" field). This was not an issue for \citet{elmahdy_privacy_2022} because they attacked canaries, whose structure is exactly known. However, for real data, we deem it necessary to initially filter plausible tokens with a MLM head.

Moreover, a forward-pass is executed for each candidate token during step 2. As such, selecting only a reasonable number of candidates significantly accelerates the attack.

\begin{algorithm}[t!]
\caption{Computing and maximizing $P_\theta^\text{tok}$}\label{alg:token_level_proxy}
\algrenewcommand\algorithmicprocedure{\textbf{procedure}}
\begin{algorithmic}[1]
\Procedure{$\argmax_{v_t \in \mathcal{V}}\ P_\theta^\text{tok}$}{$v_t \given \widetilde{d}, y, v_1, \dots, v_{t-1}$}
    \Statex \CommentLine{Getting candidates from auxiliary public MLM}
    \State $[\ g_i\ ]_{1 \le i \le |\mathcal{V}|} \gets \text{MLM}\PUB(\widetilde{d}, v_1, \dots, v_{t-1})$
    \State $[\ c_i\ ]_{1 \le i \le N_c} \gets \text{TOP}(N_c, [\ g_i\ ]_{1 \le i \le |\mathcal{V}|}).\text{indices}$
    \State $[\ g_i\ ]_{1 \le i \le N_c} \gets \text{TOP}(N_c, [\ g_i\ ]_{1 \le i \le |\mathcal{V}|}).\text{values}$
    \Statex \CommentLine{Computing target model loss}
    \State $[\ l_i\ ]_{1 \le i \le N_c} \gets [\ \text{LOSS}(f_\theta(c_i, y))\ ]_{1 \le i \le N_c}$
    \Statex \CommentLine{Converting into likelihoods}
    \State $[\ \widehat{g}_i\ ]_{1 \le i \le N_c} \gets \text{SOFTMAX}([\ g_i\ ]_{1 \le i \le N_c})$
    \State $[\ \widetilde{l}_i\ ]_{1 \le i \le N_c} \gets 2 - [\ l_i\ ]_i / \text{MEDIAN}([\ l_i\ ]_i)$
    \State $[\ \widehat{l}_i\ ]_{1 \le i \le N_c} \gets \text{SOFTMAX}([\ \widetilde{l}_i\ ]_{1 \le i \le N_c})$
    \Statex \CommentLine{Aggregating}
    \State $[\ p_i\ ]_{1 \le i \le N_c} \gets [\ \text{MEAN}(\widehat{l}_i, \widehat{g}_i)\ ]_{1 \le i \le N_c}$
    \Statex \CommentLine{Sampling with the final likelihoods}
    \State $v_t \gets \text{SAMPLE\_ONE}([\ p_i\ ]_{1 \le i \le N_c})$
\EndProcedure
\end{algorithmic}
\end{algorithm}

\paragraph{Details on step 2 (line 5 in algo. \ref{alg:token_level_proxy}, block B in fig. \ref{fig:proxy_optimization})}

For each of the $N_c$ candidate tokens, we execute a single forward-pass on the document, wherein the target token is replaced by the candidate token. These forward-passes can be arranged into one or several mini-batches depending on the available graphical memory. The loss applied is identical to the one used during training, which is the cross-entropy loss. However, we did not average over the mini-batch samples: $l_i$ corresponds to the loss for candidate token $c_i$, and to it alone.

The underlying idea is that the target model is trained to minimize the loss during training. Therefore, we assume that the loss is especially low for the specific token that was processed during training.

\paragraph{Details on step 3 (line 6 in algo. \ref{alg:token_level_proxy}, block C in fig. \ref{fig:proxy_optimization})} This step involves converting the PUB-MLM logits into likelihoods using a softmax function. We incorporated two heuristics: a temperature parameter \cite{ficler_controlling_2017, ackley_learning_1985}, following the approaches of \cite{carlini_extracting_2021, yu_bag_2023, lee_language_2022}, and a decaying temperature as in \cite{carlini_extracting_2021}, to promote the exploration of unexpected tokens only for the first tokens of the reconstruction.

\paragraph{Details on step 4 (lines 7--8 in algo. \ref{alg:token_level_proxy}, block D in fig. \ref{fig:proxy_optimization})} This step involves the conversion of the losses of the target model into likelihoods. We must consider that:

\begin{itemize}
    \item We want to minimize the loss (unlike the logits).
    \item We observed that the output loss usually consists in a majority of small values (e.g. of the order of $10^{-5}$), and some much larger outliers (e.g. of the order of $10^0$).
\end{itemize}

We used equation \ref{eq:loss_to_likelihood}. First, losses are normalized by their median, attenuating the influence of the outliers. Then, we subtract the resulting value from $2$ to position the majority of values around $1$. Ultimately, a softmax function with decaying temperature is applied, as in step 3.

\begin{equation} \label{eq:loss_to_likelihood}
    [\ \widehat{l} \ ]_{1 \le k \le n} = \underset{1 \le j \le N_c}{\text{SOFTMAX}} \left( 2 - \frac{l_j}{\text{MEDIAN}([\ l_i\ ])_{1\le i\le N_c}}\right)
\end{equation}

\paragraph{Details on step 5 (line 9 in algo. \ref{alg:token_level_proxy}, block E in fig. \ref{fig:proxy_optimization})}

In this step, we integrate the PUB-MLM likelihoods with those of the target model. We explored two possibilities: an arithmetic weighted mean and a geometric weighted mean. Each of them offers distinct advantages. The arithmetic mean guarantees that both likelihoods always have the same relative importance, while the geometric means ensures that if one of the two likelihoods is exceptionally low, the mean will approach zero regardless of the other's value. In practice, we opted for the geometric mean due to its superior results.

\paragraph{Details on step 6 (line 10 in algo. \ref{alg:token_level_proxy}, block F in fig. \ref{fig:proxy_optimization})}

Given the final likelihoods of the tokens, the remaining step consists of sampling one of them. This is a classical task in text generation \cite{holtzman_curious_2019}. Consistent with the results of \cite{yu_bag_2023, lee_language_2022}, we utilize top-$p$ sampling (also known as nucleus-$\eta$ \cite{holtzman_curious_2019}). The selection of parameter $p$ significantly influences the reconstructed field, and must be chosen in harmony with the temperature of the softmax functions.

\subsection{Variants of the attack} \label{sec:variants}

\paragraph{Targeting a MLM head}

As explained in section \ref{sec:background_doc_models}, we also target models trained on the MLM task, referred to as \textit{private} MLM, unlike the PUB-MLM head which is used to optimize the token-level proxy (see section \ref{sec:token_level_optim}).

The attack method differs slightly when targeting such models, because we do not need to use another MLM as an auxiliary model. For steps 1 and 3, we replace the PUB-MLM head with the \textit{private} MLM under attack. Then, we skip steps 2, 4 and 5, and directly utilize the likelihoods of the target model to sample the next token.

\paragraph{The multi-shot variant} For this variant, a membership inference metric is employed to sort the fields by plausibility of their reconstructions. In real-life scenarios, the adversary would set a threshold and only keep the most plausible ones.

Various membership inference metrics have been implemented in NLP \cite{carlini_extracting_2021, lee_language_2022, yu_bag_2023, lehman_does_2021, mireshghallah_quantifying_2022}. Most of them rely on comparing the \textit{likelihood} of a token or a sentence with respect to the target model and with respect to another model. Yet, computing the likelihood of a sentence for encoder-only models remains computationally intensive \cite{mireshghallah_quantifying_2022, ishihara_training_2023}. Therefore we use the same approximation as for the reconstruction: computing the field likelihoods in an autoregressive manner:

\begin{itemize}
    \item For each token $v_t$ in a field $\widetilde{f}$, we define its \textit{target likelihood} as the probability $p_i$ it had during sampling at line 10 in algorithm \ref{alg:token_level_proxy}. We also define its PUB-MLM likelihood as $\widehat{g_i}$, computed with the auxiliary MLM.
    \item We define the likelihood of a field as the product of the likelihoods of its tokens. We also use these token-level likelihoods to define the \textit{perplexity} of a field \cite{jelinek_perplexitymeasure_1977}.
\end{itemize}

While these definitions are approximations for encoder-only models, they offer the benefit of directly employing the PUB-MLM likelihood $\widehat{g_i}$ which is already computed during the attack. This circumvents the need for separate likelihood computation with an auxiliary model, as required in \cite{mireshghallah_quantifying_2022, carlini_extracting_2021, yu_bag_2023}, except when targeting a private MLM head (see above). The five membership inference metrics we implement are:

\begin{itemize}
    \item \textit{Raw perplexity}: the PUB-MLM perplexity.
    \item \textit{Perplexity ratio}: the ratio between the PUB-MLM perplexity and the target perplexity.
    \item \textit{Raw and ratio}: the product of the two metrics above, aiming at prioritizing fields that have high PUB-MLM perplexity along with a high perplexity ratio.
    \item \textit{Max. token likelihood gap}: the maximum difference between the private likelihood and the PUB-MLM one for every token (inspired from "high confidence" in \cite{yu_bag_2023}).
    \item \textit{Max. token likelihood ratio}: similar to above, replacing the difference with a ratio.
\end{itemize}

\section{Evaluation protocol} \label{sec:evaluation_protocol}

\subsection{Datasets and training conditions} \label{sec:dataset_training_conditions}

\begin{table}[b!]
\begin{center}
\begin{tabular}{|@{\hspace{5pt}}c@{\hspace{5pt}}c@{\hspace{5pt}}|c@{\hspace{5pt}}c|c@{\hspace{5pt}}c|}
\hline
\multirow{2}{*}{\textbf{Dataset}} & \multirow{2}{*}{\textbf{Task}} & \multicolumn{2}{c|}{\textbf{Acc LayoutLM}} & \multicolumn{2}{c|}{\textbf{Acc BROS}}    \\
        &        & \textbf{Unimod}   & \textbf{Bimod}    & \textbf{Unimod}   & \textbf{Bimod} \\ \hline
FUNSD   & MLM    & 0.545  & 0.545 & 0.546  & 0.546 \\
SROIE   & MLM    & 0.540  & 0.538 & 0.541  & 0.539 \\
FUNSD   & EE-BIO & 0.762  & 0.746 & 0.820  & 0.817 \\
SROIE   & EE-BIO & 0.960  & 0.967 & 0.969  & 0.959 \\
FUNSD   & EE-SPD & 0.760  & 0.760 & 0.815  & 0.804 \\
SROIE   & EE-SPD & 0.952  & 0.935 & 0.965  & 0.965 \\
FUNSD   & EL     & 0.240  & 0.193 & 0.000  & 0.202 \\ \hline
\end{tabular}
\caption{The best validation accuracy of our models. For each of our 7 tuples (dataset, task), we trained unimodal and bimodal models with either LayoutLM or BROS backbone. These results are consistent with the one announced in \cite{hong_bros_2021}. For each of these 28 settings, we attacked both the model with the best validation accuracy (Precision criterion, presented here) and with the lowest validation loss (Loss criterion, presented in the appendix, see table \ref{tab:detailed_results}).}
\vspace{-11pt}
\label{tab:models_perfs}
\end{center}
\end{table}

\paragraph{Datasets}

We evaluate the CDMI attack on models trained on two datasets: FUNSD \cite{jaume_funsd_2019} and SROIE \cite{huang_icdar2019_2019}, which are popular benchmarks for KIE tasks (see section \ref{sec:background_doc_models}). FUNSD consists of 199 scanned forms from the tobacco industry and includes annotations for both EE and EL tasks. We attack the fields annotated as "Anwers". For its part, SROIE incorporates 626 scanned receipts, only annotated for EE task. We attack the fields containing the date, company, address, and total amount. For both datasets, we only retain fields comprising between 3 and 15 tokens to disregard those that are either meaningless or too computationally demanding.

Each of our datasets is randomly partitioned into three non-overlapping parts (see appendix for details):

\begin{itemize}
    \item A validation set, used to evaluate the generalization performance of models during their training phase.
    \item A private train set, comprising half of the remaining documents, and on which the models we target are trained.
    \item A public train set, containing the remaining documents, used to train auxiliary models.
\end{itemize}

\paragraph{Training conditions} We trained models in 28 different settings. For each of the 7 tuples (dataset, task) presented in table \ref{tab:models_perfs},  we trained unimodal and bimodal models with either LayoutLM or BROS backbone. For each configuration, we trained our models for many epochs (300 for MLM tasks, 150 for the others), and saved the models at each epoch. Ultimately, we select the epoch we target with respect to one of the following \textit{criteria}:

\begin{itemize}
    \item Precision: the epoch with the best validation accuracy.
    \item Loss: the epoch with the best validation loss.
\end{itemize}

Using these 28 configurations and these 2 criteria, we evaluated both the one-shot and the multi-shot variant of our attack. This results in \underline{112 different attack scenarios}.

\paragraph{Hyperparameters: transferring from LayoutLM to BROS} The CDMI attack described in section \ref{sec:cdmi_method} involves several hyperparameters: the number of candidates $N_c$ in step 1, the temperature parameter and decay in steps 3 and 4, the averaging method and its weight in step 5, the parameter $p$ for top-$p$ sampling in step 6, and the number of attempts $n_a$ for the multi-shot variant. We tuned these hyperparameters for the 14 configurations with LayoutLM backbone and Precision criterion, optimizing the evaluation metrics of the one-shot variant (see section \ref{sec:eval_oneshot} and the appendix).

Finally, we used each of these 14 sets of hyperparameters for 8 attacks: one-shot or multi-shot, Precision or Loss criterion, and LayoutLM or BROS backbone, yielding in a total of 112 evaluation settings. Given that the hyperparameters were calibrated with LayoutLM backbone, we anticipate the accuracy of CDMI to be superior with LayoutLM than with BROS. These enable us to assess the \textit{transferability} of our attack, as we measure its performance on an architecture for which it was not optimized, namely BROS.

\paragraph{Hardware and computation time} Each of these 112 evaluations was performed on a fraction of 0.1 Nvidia A100 80GB. The mean computation time was 20:40 hours for each experiment, resulting in a total of about 2300 hours.

\subsection{Evaluating the one-shot variant} \label{sec:eval_oneshot}

\paragraph{Four evaluation metrics} To evaluate the quality of the one-shot attack, we compare each reconstruction $\widetilde{f}$ to its corresponding ground truth $f$. We employed four metrics for this purpose: the Perfect Reconstruction $\text{PR} = \mathbb{1}(f = \widetilde{f})$, the Hamming Distance (HD) \cite{hamming_error_1950}, the normalized Levenshtein Distance (LD) \cite{levenshtein_binary_1966}, and the normalized Jaro-Winkler Distance (JWD) \cite{jaro_advances_1989, winkler_string_1990}. We computed these metrics using the tokens of the fields, ignoring their alphanumeric decoding by the tokenizer, and we averaged them across all fields.

\paragraph{Our baseline} A logical analysis of the other fields and headers can often enable guessing many fields within a document. Consequently, we compare our one-shot attack to a baseline which represents the reconstruction that are feasible with public information only. In practice, this baseline corresponds to the reconstruction of the fields using only the PUB-MLM likelihoods in figure \ref{fig:proxy_optimization}.

We compute the four evaluation metrics with this baseline, and compare them to our attack. In this way, we assess how useful it is to have access to the target model to obtain accurate reconstructions. Using an idea similar to the \textit{advantage} often used in cryptography, we introduce the \textit{improvement factor} (IpF), which represents the mean improvement between the attack and the baseline for our four metrics. For example, $\text{IpF} = 1.5$ means that the attack is 50\% better than the baseline on average. The improvement factor is defined in equation \ref{eq:improvement_factor}, where "att" represents the attack, and "base" the baseline. We also add a parameter $\varepsilon$ to bound the improvement factor.

\begin{equation} \label{eq:improvement_factor}
\text{IpF} = \frac{
    \frac{\text{PR}\ATT+\varepsilon}{\text{PR}\BASE+\varepsilon}
    + \frac{\text{HD}\BASE+\varepsilon}{\text{HD}\ATT+\varepsilon}
    + \frac{\text{LD}\BASE+\varepsilon}{\text{LD}\ATT+\varepsilon}
    + \frac{\text{JWD}\BASE+\varepsilon}{\text{JWD}\ATT+\varepsilon}
}{4}
\end{equation}

\subsection{Evaluating the multi-shot variant} \label{sec:eval_multishot}

The evaluation of the multi-shot variant must account for both the order of the reconstructions with respect to the membership inference metric ($F$ and $\widetilde{F}$), and the similarity between the fields and their reconstructions ($f_i \in F$  vs. $\widetilde{f_i} \in \widetilde{F}$). Thus, our evaluation metrics should satisfy the two following mathematical properties: 

\begin{enumerate}
    \item With fixed values $f_i$ and $\widetilde{f_i}$ for each $i$, the metric is maximized when $F$ and $\widetilde{F}$ are sorted in descending order of similarity between $f_i$ and $\widetilde{f_i}$.
    \item With a fixed order of $F$ and $\widetilde{F}$, for each $i$, the metric increases when the similarity between $f_i$ and $\widetilde{f_i}$ increases.
\end{enumerate}

\begin{figure}[t!]
    \centering
    \includegraphics[width=0.99\columnwidth]{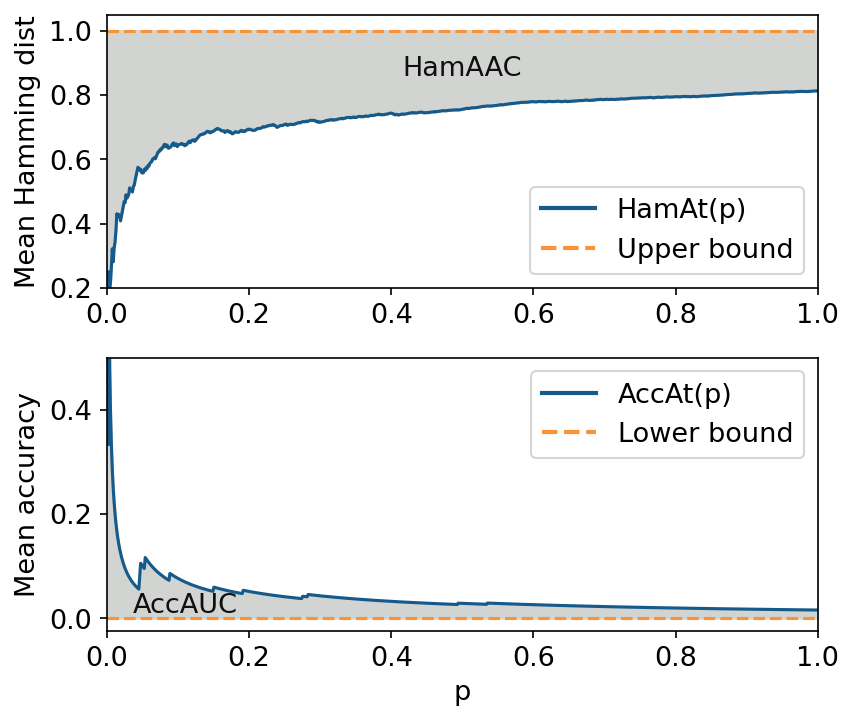}
    \caption{$AccAUC$ and $HamAAC$ computation examples. Acc($p$) denotes the mean accuracy of the top-$p$ fields the adversary is the most confident in. The greater it is, the more accurate the reconstructions are. A peaky and decreasing shape means that the membership inference metric accurately sorts the reconstruction attempts. This is why we seek to maximize its Area Under the Curve ($AccAUC$). Idem for the Area Above the Curve with the Hamming distance ($HamAAC$).}
    \label{fig:hamaac_acauc_example}
\end{figure}

However, common metrics used in the literature do not satisfy these properties. Indeed, existing work such as \cite{carlini_extracting_2021, yu_bag_2023} assess separately the quality of the reconstructions and the membership inference attack. For example, with the regular AUC score computed with the ROC curve, a value of 1.00 is achieved in the two following situations, despite the apparent superiority of the latter: (1) a single perfect reconstruction is ranked highest, trailed by 99 inaccurate reconstructions; and (2) all 100 reconstructions are perfect. This is because AUC score ignores the \textit{prevalence} of perfect reconstruction, which is necessary to meet the second property we want.

This motivates the introduction of two new metrics that satisfy these properties. Figure \ref{fig:hamaac_acauc_example} demonstrates an example of their computation. We will introduce them briefly, and we refer to the appendix for rigorous definitions and proofs that they satisfy the two desired properties.

\paragraph{Accuracy-AUC and Hamming-AAC} First, we introduce the Accuracy-at-$p$ metric for $p \in [0, 1]$ ($AccAt(p)$). It represents the accuracy of the reconstruction within the top-$p$ fields the adversary is the most confident in, i.e. the one with the greatest membership inference metric. Then, we plot $AccAt(p)$ versus $p$. A greater Area Under the Curve ("AUC") means that that the reconstruction are more accurate. This motivates the introduction of the Accuracy-AUC metric.

Similarly, $HamAt(p)$ represents the mean normalized Hamming distance \cite{hamming_error_1950} within the top-$p$ fields the adversary is the most confident in. However, we look at the Area Above the Curve ("AAC") because it is a distance and not a similarity. This results in the Hamming-AAC metric. Both the $AccAUC$ and $HamAAC$ metrics take values between 0 and 1, and increase as the quality of the attack improves.

\section{Experimental results} \label{sec:experimental_results}

This section presents the experimental results of our attack and discusses our main findings. For a comprehensive presentation of our results, please refer to the appendix. Our experiments show that CDMI achieves robust results, allowing for meaningful reconstructions with every task and dataset we evaluated, including numbers, dates, and names up to 8 tokens. Table \ref{tab:example_perf_rec} shows examples of perfect reconstruction.

In our experiments, the maximum value of $AccAt(1.0)$ that we reach is 0.041, meaning that under optimal conditions, the adversary can flawlessly reconstruct \underline{4.1\% of the fields.} For comparison, under identical conditions, our baseline only recovers 0.77\% of the fields. Furthermore, the maximum $AccAt(0.05)$ registered is a promising \underline{22.5\%}. This indicates that in this scenario, when the adversary employs a membership inference metric to retain the 40 fields they are most confident in (5\% of the dataset), they can attain up to 9 perfect reconstructions. This performance far exceeds our baseline's score at 2.5\%, corresponding to just a single field.

\begin{table}[b!]
\centering
\begin{tabular}{c@{\hspace{5pt}}c@{\hspace{5pt}}c@{\hspace{5pt}}c@{\hspace{5pt}}c@{\hspace{5pt}}c}
    \hline
    \textbf{Data}  & \textbf{Archi} & \textbf{Task} &\textbf{Reconstruction} & \textbf{Len} & \textbf{Occ}  \\ \hline
    SRO & LayoutLM & EE-SPD & \makecell{restaurant jiawei\\[-4pt] jiawei house} & 6 & 1 \\
    SRO & LayoutLM & EE-BIO & \makecell{guardian health and\\[-4pt] beauty sdn bhd} & 8 & 1 \\
    FUN & LayoutLM & EL & m. a. peterson & 5 & 1 \\ 
    FUN & LayoutLM & EE-SPD & r. g. ryan  & 5 & 1 \\ 
    SRO & LayoutLM & MLM & \makecell{lim seng tho\\[-4pt] hardware trading} & 6 & 1 \\ 
    SRO & LayoutLM & MLM & 101. 75 & 3 & 2 \\ 
    FUN & LayoutLM & MLM & april 13, 1984 & 4 & 1 \\
    FUN & BROS & MLM & 1. 500. 00 & 6 & 1 \\
    FUN & BROS & MLM & dr. a. w. spears & 7 & 1 \\ \hline
\end{tabular}
\caption{Examples of perfect reconstructions.}
\label{tab:example_perf_rec}
\end{table}

\subsection{Performance of the attack} \label{sec:experimental_perfs}

\paragraph{Factors influencing the one-shot variant with LayoutLM} We observe that certain configurations result in significant outcomes. For example, when selecting the checkpoint with the best validation precision from a bimodal MLM trained on FUNSD, we achieve an improvement factor of 1.395. This means that the CDMI reconstructions are approximately 40\% more effective than the ones of our baseline (see line 2 in table \ref{tab:detailed_results} in the appendix for more details).

\begin{figure}[t!]
    \centering
    \includegraphics[width=\columnwidth]{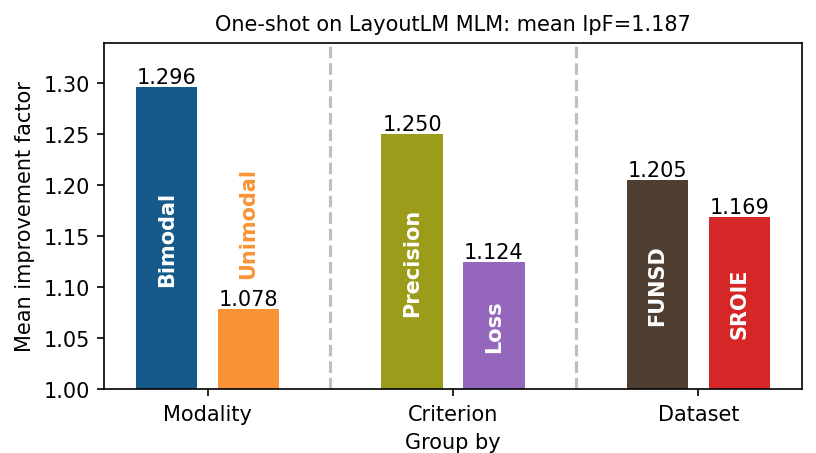}
    \caption{Factors influencing the performance of the one-shot attack on a LayoutLM with MLM task. We attack 8 different models (2 modality, 2 criteria, 2 datasets) with an average improvement factor of $\text{IpF} = 1.187$. Among them, the 4 attacks implying a bimodal model are more accurate than those against a unimodal model (IpF of $1.296$ vs. $1.078$). Similarly, the attacks are more accurate with the Precision criterion, and with the FUNSD dataset.}
    \label{fig:mlm_oneshot_factors}
\end{figure}

However, some models are more vulnerable than others. Figure \ref{fig:mlm_oneshot_factors} compares the performance of the one-shot attack for the 8 MLM models we trained with LayoutLM backbone: unimodal or bimodal, Precision or Loss criterion, and FUNSD or SROIE dataset (see lines 1-8 in table \ref{tab:detailed_results} for more details). Interestingly, the difference between FUNSD and SROIE datasets is minor, showing that our attack is robust to at least two type of documents. Moreover, we observe that Precision criterion leads to more vulnerable models than Loss criterion. This is not surprising because they are trained for more epochs, so the model had more time to memorize its training data. Finally, we observe that bimodal models are much more vulnerable than unimodal ones, with a mean improvement factor of 1.296 compared to 1.078. This indicates that the visual encoder memorizes information about the training sample, which our attack efficiently extracts. These aspects are discussed further in section \ref{sec:ablation_studies}.

For non-MLM tasks, the factors influencing the one-shot attack are different. Indeed, the attack is very efficient on SROIE dataset, with a mean improvement factor of 1.224. However, it obtains poor results on FUNSD, with a mean improvement factor of 1.036, indicating that the attack is not significantly better than the baseline. We explain this result by observing that these models exhibit significantly lower accuracy on FUNSD compared to SROIE, so it is not surprising that the attack can extract less information from them (see columns 6-7, lines 9-24 in table \ref{tab:detailed_results} for more details).

\paragraph{Comparison between the tasks}

\begin{figure}[b!]
    \centering
    \includegraphics[width=\columnwidth]{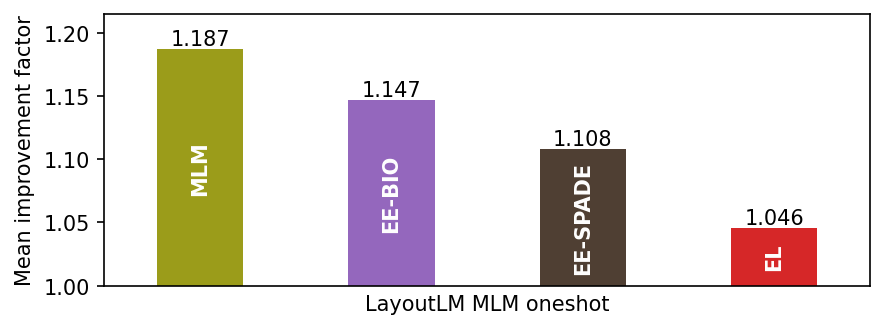}
    \includegraphics[width=\columnwidth]{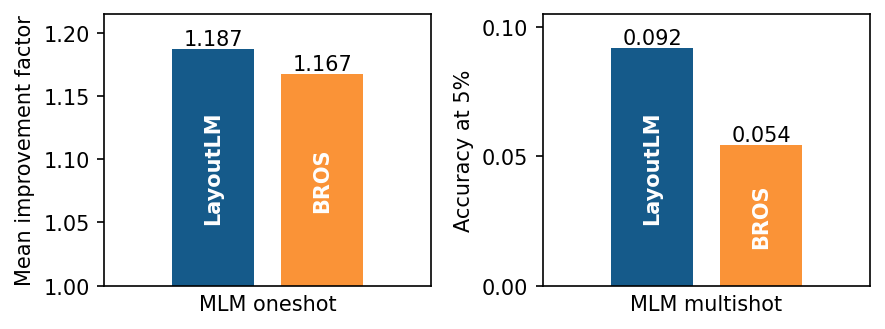}
    \caption{Performance comparison based on the backbone or task. The top graph shows the average performance of the attack with LayoutLM backbone, for the four possible tasks. The bottom graphs compare the average performance of the attack on the MLM models with LayoutLM backbone (in blue) or BROS backbone (in orange), for both the one-shot variant (left) and the multi-shot variant (right).}
    \label{fig:tasks_backbone_comparison}
\end{figure}

With LayoutLM backbone, some tasks are easier to attack than others, as outlined in figure \ref{fig:tasks_backbone_comparison}. MLM task is the most vulnerable, as expected. Indeed, it is designed to reconstruct masked tokens, aligning with the objective of the reconstruction. On the opposite, models trained with EL task exhibit lower validation accuracy compared to EE-BIO and EE-SPADE, which explains their low susceptibility to our attack. Since the model struggles in learning meaningful patterns to solve the task, using its loss as per the CDMI approach makes it less efficient.

However, we observe that for every task and dataset, there exists a configuration yielding an improvement factor of a minimum of $1.10$. This suggests that our reconstruction attack could be generalized to a wider range of fine-tuning tasks, such as document classification and question answering.

\paragraph{Comparison between LayoutLM and BROS} As detailed in section \ref{sec:evaluation_protocol}, we have optimized our attack with the LayoutLM backbone, and then evaluated its transferability to BROS architecture. Thus, it is unsurprising that we obtain stronger results for LayoutLM than BROS. For example, we obtain poor results for most one-shot and multi-shot configurations with non-MLM tasks when using BROS backbone. 

However, the outcomes of CDMI for MLM tasks on BROS are highly competitive compared to those on LayoutLM, as outlined in figure \ref{fig:tasks_backbone_comparison}. For the one-shot variant, there is a minor gap between LayoutLM and BROS backbones, with a mean improvement factor of $1.187$ and $1.167$, respectively. Our best performance for the one-shot variant is even obtained with BROS backbone, with an impressive improvement factor of 1.54. For the multi-shot variant with BROS backbone, 5.4\% of the fields are perfectly reconstructed on average within the top-5\% fields (Accuracy-at-5\%). This is much higher than the accuracy of 1.3\% obtained with the baseline, showing that the model memorizes a significant part of its training set.

The successful transferability of our attack to BROS backbone in some settings suggests that it could likely be adapted to other document understanding model architectures.

\paragraph{The multi-shot variant} 

We observe that many configuration yielding good results with the one-shot variant perform poorly with the multi-shot variant. Thus, in these configurations, even though the model memorizes information about its training data, it is not sufficient to perfectly reconstruct a significant portion of the fields with high confidence.

\begin{figure}[t!]
    \centering
    \includegraphics[width=\columnwidth]{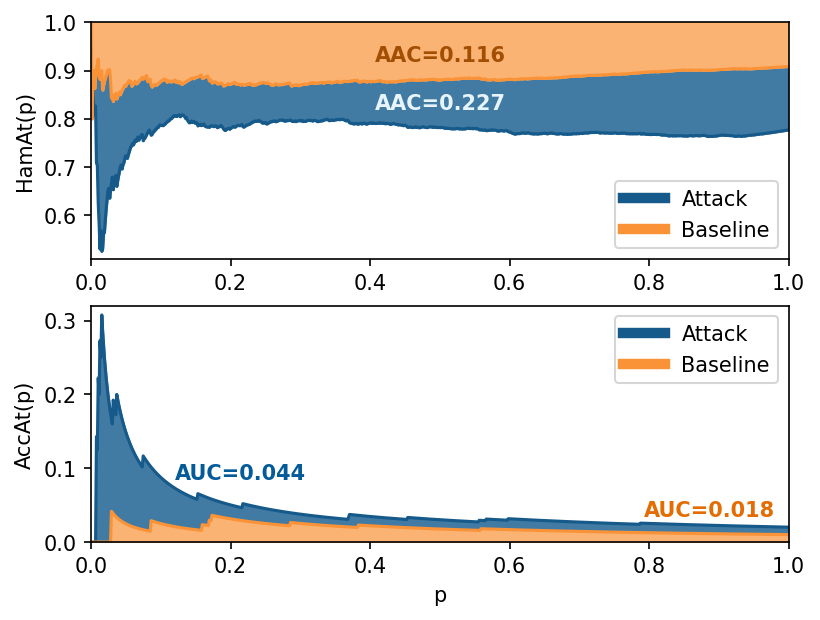}
    \includegraphics[width=0.99\columnwidth]{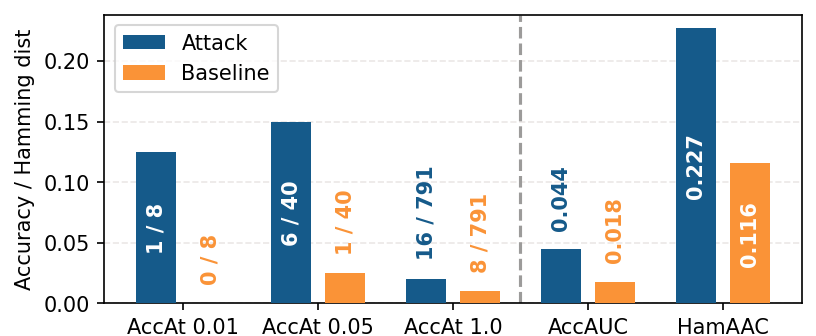}
    \caption{Comparison between the attack and the baseline in a multi-shot setting, with a bimodal model, LayoutLM backbone, EE-SPADE task, SROIE dataset and Loss criterion. On the upper plot, the Hamming-at-$p$ distance is much lower for the attack. On the middle plot, the Accuracy-at-$p$ of the attack is decreasing, so the membership inference metric accurately sorts the most credible reconstructions. On the last plot, the evaluation metrics are higher for the attack: for instance, with $p = 0.05$, the attack perfectly reconstructs $6/40 = 15\%$ of the fields, vs. $1/40 = 2.5\%$ for the baseline.}
    \label{fig:non_mlm_multishot}
\end{figure}

However, some configurations lead to strong results for both MLM and non-MLM tasks. For example, figure \ref{fig:non_mlm_multishot} displays a successful attack on a very realistic model trained on EE-SPADE task, which is targeted at epoch 4, when its validation loss is minimal. The adversary perfectly reconstructs 15\% of the fields they are most confident in. Moreover, the peaked shapes of $AccAt(p)$ and $HamAt(p)$ indicate that the membership inference metric efficiently sorts the good reconstructions first. As a result, CDMI reconstructions considerably outperform our baseline in all evaluation metrics.

This proves that even non-generative models trained on entity extraction tasks can be successfully attacked.

\subsection{Ablation studies} \label{sec:ablation_studies}

In this section, we demonstrate that neither overfitting nor data duplication accounts for the performance of our attack. Moreover, we demonstrate that we can extract information memorized by the visual modality, proving that documents should be considered as a distinct data type, susceptible to attacks that can exploit their bimodal nature.

\begin{figure}[b!]
    \centering
    \includegraphics[width=0.99\columnwidth]{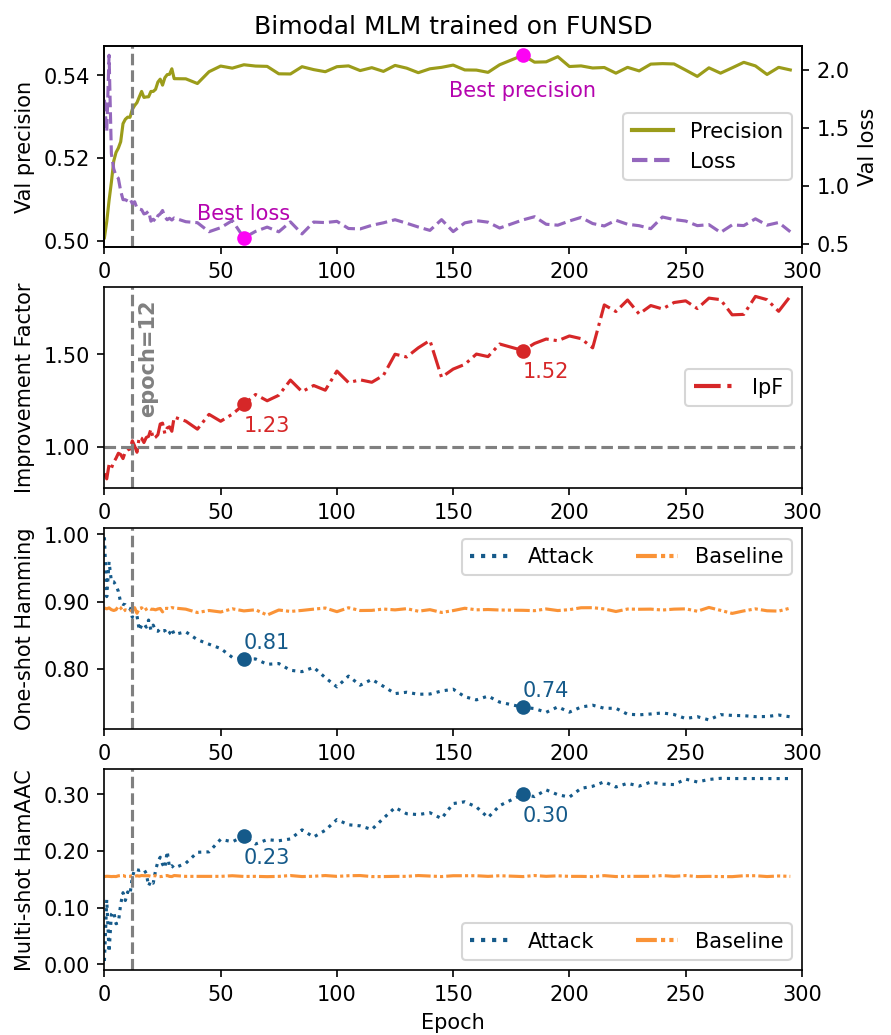}
    \caption{Attacking a bimodal LayoutLM model trained on FUNSD with MLM task and $N_c = 256$. The first graph plots the validation loss and accuracy, with the best values achieved at epoch 60 and epoch 180, respectively. The two following plots display the performance of the one-shot attack, and the last one shows the performance of the multi-shot attack. In both cases, the attack is more accurate than the baseline from epoch 12, well before the best validation loss.}
    \label{fig:mlm_vs_epoch}
\end{figure}

\paragraph{Overfitting} As detailed in section \ref{sec:background_privacy_attacks}, overfitting often favors the memorization of training data, even though it is not a requisite condition \cite{yeom_privacy_2018, zhang_understanding_2021, feldman_does_2020, carlini_extracting_2021}. We also observed this phenomenon with CDMI. From the early stages of training, even before the model can overfit, training data are memorized. And the attack performance gradually increases over epochs, even when validation accuracy starts to decrease.

To verify the correlation, we executed the experiment depicted in figure \ref{fig:mlm_vs_epoch}. We evaluate the one-shot and multi-shot CDMI attacks on 40\% of the fields for one out of 5 model checkpoints between epoch 1 and 300. First, we observe that overfitting is evident, as the best validation loss is achieved much earlier than the best validation precision. However, the attack outperforms the baseline quite early, after just 12 epochs, showing that the model memorizes training data without needing to overfit. Finally, all the performance metrics of our attack increase with the number of training epochs, and persist beyond the epoch with the optimal precision.

This shows that our attack is efficient from the early epochs of training, and that simple regularization techniques are unlikely to be sufficient to ensure the security of the data used to train document understanding models.

\paragraph{Data duplication} Data duplication is known to favor memorization of training data (see section \ref{sec:background_privacy_attacks} and \cite{kandpal_deduplicating_2022}). Certain definitions such as $k$-eidetic memorization \cite{carlini_extracting_2021} even include the number of duplicates. However, defining duplication for document data is not straightforward. In our case, we seek to reconstruct the fields in their context: for example, the reconstruction of "24.8 mm" only becomes useful when it is linked to the form it comes from. Therefore, we deemed that two forms with identical headers but different answers were not duplicates, contrary to shifted or rotated documents.

We did not implement an automatic count of duplication within our datasets; however we manually checked the number of occurrences for a few documents (see "Occ" column in table \ref{tab:example_perf_rec}). Even though we observed some duplicates in the SROIE dataset, we did not detect any of them in FUNSD. Thus, we conclude that the good results of our attack is presumably not linked to data duplication. However, further investigating this issue would be a promising topic for future research, as numerous document datasets contain a great number of similar documents (e.g. ID cards, etc.).

\begin{figure}[t!]
    \centering
    \includegraphics[width=\columnwidth]{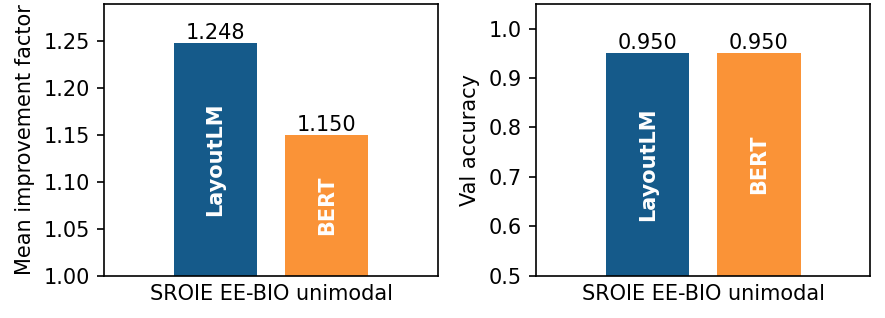}
    \includegraphics[width=\columnwidth]{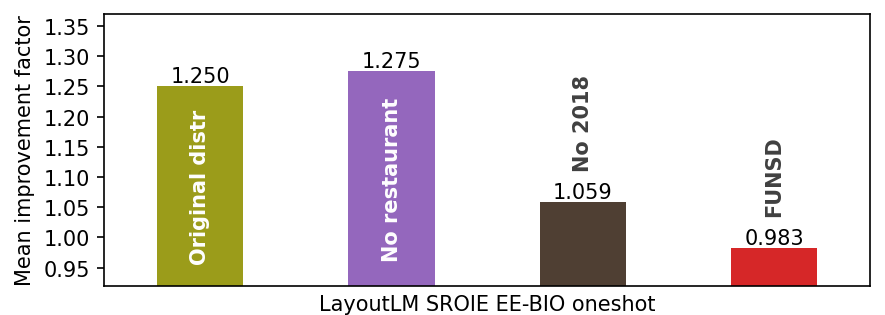}
    \caption{Upper plot: our attack performs better on LayoutLM backbones than BERT. This confirms that layout contributes to memorization, making document models more vulnerable. Lower plot: the performance of attacks using auxiliary public MLM trained on four datasets with increasing distribution shift. Our attack remains feasible if the adversary has access to different yet similar data distribution.}
    \label{fig:ablation_study}
\end{figure}

\paragraph{Importance of layout and visual modality} Our attack was designed to efficiently reconstruct data from layout-aware document models. However, it could be possible that it only extracts information from textual content, and would be just as effective on text-only models. We conducted the following experiments to demonstrate that it is not the case, and that both layout and visual modality contribute to memorization.

As shown in the upper plot of figure \ref{fig:ablation_study}, we trained several models with BERT backbone \cite{devlin_bert_2019}, which does not take the layout into account, and compared them with those using LayoutLM backbone. We specifically used models trained on SROIE with the EE-BIO task, as in this setting, the validation accuracy between BERT and LayoutLM is similar. This means that a difference in memorization cannot be attributed to a difference in accuracy. We observed that the attack is much more successful with LayoutLM, with a mean improvement factor of 1.248 compared to 1.150 with BERT. This proves that layout information contributes to memorization, making the layout-aware models more vulnerable than their textual counterparts.

The significance of visual modality was already discussed in section \ref{sec:experimental_perfs}. As depicted in figure \ref{fig:mlm_oneshot_factors}, bimodal models are more susceptible to our attack than unimodal ones. To confirm that the information is memorized within visual modality, we conducted an experiment using bimodal models trained on MLM task with LayoutLM backbone. We obtained a mean improvement factor of 1.296 when attacking these models normally. However, when we deactivated the visual modality by replacing it with Gaussian noise in these models, the mean improvement factor dropped significantly to 1.044. This confirms that our attack extracts information that is memorized within the visual modality.

\paragraph{Robustness to distribution shifts} As explained in section~\ref{sec:threat_model}, we assume that the adversary knows the target distribution $\mathcal{D}$, and uses it to train an efficient auxiliary public MLM. This hypothesis is plausible given the accessibility of many datasets online, but it gives an advantage to the adversary. 

To evaluate its impact, we conducted attacks on SROIE dataset using auxiliary public MLM trained on four datasets with increasing distribution shifts (see figure \ref{fig:ablation_study}). The first one was trained with the original distribution. For the second one, we removed all restaurant receipts from SROIE, which represents a sub-population of about 10\%. We obtained a similar improvement factor, indicating that our attack remains feasible with a minor distribution shift. The third one was trained without any receipt dated 2018 (around 50\% of the dataset). This is an important distribution shift because the ‘date’ field is one of those we are attacking. The performance dropped, because the adversary is unable to accurately select relevant candidate tokens. Finally, the last auxiliary model was trained on FUNSD. Here, the distribution shift is too important and the attack is ineffective, yielding an improvement factor close to 1. We conclude that the ability of the adversary to train a well-performing auxiliary model on a nearby distribution is an important assumption for the success of our attack.

\section{Conclusion} \label{sec:conclusion}

\paragraph{A pioneer attack} In this paper, we introduce the first reconstruction attack against document understanding models: CDMI (Combinatorial Document Model Inversion). Despite similarities with some attacks against language models, what sets CDMI apart is that it is explicitly designed to handle documents as multimodal data, making it the first to target layout-aware models. We present two variants of the attack: a one-shot version, and a multi-shot version where CDMI is combined with a membership inference attack.

We also establish a meticulous protocol to assess our attack, comparing it with reconstructions that can be made with only public information. We also introduce two new evaluation metrics, Accuracy-AUC and Hamming-AAC, designed to simultaneously evaluate the reconstruction phase and the membership inference phase in the multi-shot variant.

\paragraph{Empirical evidence} We demonstrate that models trained on Key Information Extraction tasks under realistic conditions are vulnerable to reconstruction attacks. Under optimal conditions, the adversary is able to perfectly reconstruct from 4.1\% of the fields to 22.5\% when combined with a membership inference attack.

We clearly observe that the Masked Language Modelling task is more vulnerable than the Key Information Extraction ones. Moreover, we demonstrate that both layout and visual modality contribute to memorization, making document models more vulnerable than their textual counterparts, and susceptible to specific attack developments.

We show that our attack's performance is not a consequence of either overfitting or data duplication. It only takes a dozen epochs to memorize training data even without duplication. Thus, further research into alternative defense mechanisms, such as differential privacy, is necessary.

\paragraph{Recommendations} We recommend three mechanisms to make the prerequisites of our attack impossible. First, avoid open-sourcing the weights of models trained on sensitive data. Second, when serving models through an API, hide your model’s confidence score and filter repetitive queries on similar data. Third, avoid sharing anonymized documents, as they can be used for reconstruction attacks.

If these three mechanisms cannot be implemented in a particular situation, we recommend exercising the utmost caution. More advanced reconstruction attacks are likely to be developed in the near future. Our attack method can be used to estimate the privacy risk of a model, and we also advocate for the use of provable privacy-preserving techniques.

We hope our work will alert researchers and practitioners to the privacy risks of document understanding models, and lay the foundation for the development of robust, privacy-preserving architectures.

\setlength\bibitemsep{0.75\itemsep}
\printbibliography

@article{dwork_algorithmic_2014,
	title = {The {Algorithmic} {Foundations} of {Differential} {Privacy}},
	doi = {10.1561/0400000042},
	journal = {FnT-TCS},
	author = {Dwork, Cynthia and Roth, Aaron},
	year = {2014},
}

@incollection{dwork_calibrating_2006,
	title = {Calibrating {Noise} to {Sensitivity} in {Private} {Data} {Analysis}},
	booktitle = {{TCC}},
	author = {Dwork, Cynthia and McSherry, Frank and Nissim, Kobbi and Smith, Adam},
	year = {2006},
	doi = {10.1007/11681878_14},
}

@article{dwork_firm_2011,
	title = {A firm foundation for private data analysis},
	doi = {10.1145/1866739.1866758},
	journal = {CACM},
	author = {Dwork, Cynthia},
	month = jan,
	year = {2011},
}

@article{domingo-ferrer_limits_2021,
	title = {The limits of differential privacy (and its misuse in data release and machine learning)},
	doi = {10.1145/3433638},
	journal = {CACM},
	author = {Domingo-Ferrer, Josep and Sánchez, David and Blanco-Justicia, Alberto},
	month = jul,
	year = {2021},
}

@inproceedings{brown_what_2022,
	title = {What {Does} it {Mean} for a {Language} {Model} to {Preserve} {Privacy}?},
	doi = {10.1145/3531146.3534642},
	booktitle = {{ACM} {FAccT}},
	author = {Brown, Hannah and Lee, Katherine and Mireshghallah, Fatemehsadat and Shokri, Reza and Tramèr, Florian},
	month = jun,
	year = {2022},
	pages = {2280--2292},
}

@article{ackley_learning_1985,
	title = {A learning algorithm for boltzmann machines},
	volume = {9},
	number = {1},
	journal = {Cognitive Science},
	author = {Ackley, D and Hinton, G and Sejnowski, T},
	month = mar,
	year = {1985},
	pages = {147--169},
}

@inproceedings{abadi_deep_2016,
	title = {Deep {Learning} with {Differential} {Privacy}},
	doi = {10.1145/2976749.2978318},
	booktitle = {{ACM} {SIGSAC} {CCS}},
	author = {Abadi, Martín and Chu, Andy and Goodfellow, Ian and McMahan, H. Brendan and Mironov, Ilya and Talwar, Kunal and Zhang, Li},
	month = jul,
	year = {2016},
	pages = {308--318},
}

@inproceedings{zhang_inference_2022,
	title = {Inference {Attacks} {Against} {Graph} {Neural} {Networks}},
	doi = {10.48550/arXiv.2110.02631},
	booktitle = {{USENIX} {Security}},
	author = {Zhang, Zhikun and Chen, Min and Backes, Michael and Shen, Yun and Zhang, Yang},
	month = aug,
	year = {2022},
}

@inproceedings{carlini_extracting_2023,
	title = {Extracting {Training} {Data} from {Diffusion} {Models}},
	doi = {10.48550/arXiv.2301.13188},
	booktitle = {{USENIX} {Security}},
	author = {Carlini, Nicholas and Hayes, Jamie and Nasr, Milad and Jagielski, Matthew and Sehwag, Vikash and Tramèr, Florian and Balle, Borja and Ippolito, Daphne and Wallace, Eric},
	month = aug,
	year = {2023},
}

@article{jelinek_perplexitymeasure_1977,
	title = {Perplexity—a measure of the difficulty of speech recognition tasks},
	volume = {62},
	doi = {10.1121/1.2016299},
	number = {S1},
	journal = {The Journal of the Acoustical Society of America},
	author = {Jelinek, F. and Mercer, R. L. and Bahl, L. R. and Baker, J. K.},
	month = dec,
	year = {1977},
	pages = {S63--S63},
}

@article{alshammari_impact_2021,
	title = {The impact of using different annotation schemes on named entity recognition},
	volume = {22},
	doi = {10.1016/j.eij.2020.10.004},
	number = {3},
	journal = {Egyptian Informatics Journal},
	author = {Alshammari, Nasser and Alanazi, Saad},
	month = sep,
	year = {2021},
	pages = {295--302},
}

@misc{subramani_survey_2021,
	title = {A {Survey} of {Deep} {Learning} {Approaches} for {OCR} and {Document} {Understanding}},
	doi = {10.48550/arXiv.2011.13534},
	author = {Subramani, Nishant and Matton, Alexandre and Greaves, Malcolm and Lam, Adrian},
	month = feb,
	year = {2021},
}

@inproceedings{kim_ocr-free_2022,
	title = {{OCR}-{Free} {Document} {Understanding} {Transformer}},
	doi = {10.48550/arXiv.2111.15664},
	booktitle = {{ECCV}},
	author = {Kim, Geewook and Hong, Teakgyu and Yim, Moonbin and Nam, JeongYeon and Park, Jinyoung and Yim, Jinyeong and Hwang, Wonseok and Yun, Sangdoo and Han, Dongyoon and Park, Seunghyun},
	year = {2022},
}

@misc{jang_categorical_2017,
	title = {Categorical {Reparameterization} with {Gumbel}-{Softmax}},
	doi = {10.48550/arXiv.1611.01144},
	author = {Jang, Eric and Gu, Shixiang and Poole, Ben},
	month = aug,
	year = {2017},
}

@article{park_cord_2019,
	title = {{CORD}: {A} consolidated receipt dataset for post-{OCR} parsing},
	url = {https://github.com/clovaai/cord},
	journal = {Document Intelligence Workshop NeurIPS},
	author = {Park, Seunghyun and Shin, Seung and Lee, Bado and Lee, Junyeop and Surh, Jaeheung and Seo, Minjoon and Lee, Hwalsuk},
	year = {2019},
}

@misc{mahloujifar_membership_2021,
	title = {Membership {Inference} on {Word} {Embedding} and {Beyond}},
	doi = {10.48550/arXiv.2106.11384},
	author = {Mahloujifar, Saeed and Inan, Huseyin A. and Chase, Melissa and Ghosh, Esha and Hasegawa, Marcello},
	month = jun,
	year = {2021},
}

@misc{zhang_text_2022,
	title = {Text {Revealer}: {Private} {Text} {Reconstruction} via {Model} {Inversion} {Attacks} against {Transformers}},
	shorttitle = {Text {Revealer}},
	doi = {10.48550/arXiv.2209.10505},
	author = {Zhang, Ruisi and Hidano, Seira and Koushanfar, Farinaz},
	month = sep,
	year = {2022},
}

@misc{yang_harnessing_2023,
	title = {Harnessing the {Power} of {LLMs} in {Practice}: {A} {Survey} on {ChatGPT} and {Beyond}},
	shorttitle = {Harnessing the {Power} of {LLMs} in {Practice}},
	doi = {10.48550/arXiv.2304.13712},
	author = {Yang, Jingfeng and Jin, Hongye and Tang, Ruixiang and Han, Xiaotian and Feng, Qizhang and Jiang, Haoming and Yin, Bing and Hu, Xia},
	month = apr,
	year = {2023},
}

@misc{wang_layout_2023,
	title = {Layout and {Task} {Aware} {Instruction} {Prompt} for {Zero}-shot {Document} {Image} {Question} {Answering}},
	doi = {10.48550/arXiv.2306.00526},
	author = {Wang, Wenjin and Li, Yunhao and Ou, Yixin and Zhang, Yin},
	month = jun,
	year = {2023},
}

@misc{tramer_considerations_2022,
	title = {Considerations for {Differentially} {Private} {Learning} with {Large}-{Scale} {Public} {Pretraining}},
	doi = {10.48550/arXiv.2212.06470},
	author = {Tramèr, Florian and Kamath, Gautam and Carlini, Nicholas},
	month = dec,
	year = {2022},
}

@misc{shao_quantifying_2023,
	title = {Quantifying {Association} {Capabilities} of {Large} {Language} {Models} and {Its} {Implications} on {Privacy} {Leakage}},
	doi = {10.48550/arXiv.2305.12707},
	author = {Shao, Hanyin and Huang, Jie and Zheng, Shen and Chang, Kevin Chen-Chuan},
	month = may,
	year = {2023},
}

@misc{ishihara_training_2023,
	title = {Training {Data} {Extraction} {From} {Pre}-trained {Language} {Models}: {A} {Survey}},
	shorttitle = {Training {Data} {Extraction} {From} {Pre}-trained {Language} {Models}},
	doi = {10.48550/arXiv.2305.16157},
	author = {Ishihara, Shotaro},
	month = may,
	year = {2023},
}

@misc{dathathri_plug_2019,
	title = {Plug and {Play} {Language} {Models}: {A} {Simple} {Approach} to {Controlled} {Text} {Generation}},
	shorttitle = {Plug and {Play} {Language} {Models}},
	doi = {10.48550/arXiv.1912.02164},
	author = {Dathathri, Sumanth and Madotto, Andrea and Lan, Janice and Hung, Jane and Frank, Eric and Molino, Piero and Yosinski, Jason and Liu, Rosanne},
	month = dec,
	year = {2019},
}

@misc{cheng_trie_2022,
	title = {{TRIE}++: {Towards} {End}-to-{End} {Information} {Extraction} from {Visually} {Rich} {Documents}},
	shorttitle = {{TRIE}++},
	doi = {10.48550/arXiv.2207.06744},
	author = {Cheng, Zhanzhan and Zhang, Peng and Li, Can and Liang, Qiao and Xu, Yunlu and Li, Pengfei and Pu, Shiliang and Niu, Yi and Wu, Fei},
	month = jul,
	year = {2022},
}

@inproceedings{carlini_extracting_2021,
	title = {Extracting {Training} {Data} from {Large} {Language} {Models}.},
	doi = {10.48550/arXiv.2012.07805},
	booktitle = {{USENIX} {Security}},
	author = {Carlini, Nicholas and Tramer, Florian and Wallace, Eric and Jagielski, Matthew and Herbert-Voss, Ariel and Lee, Katherine and Roberts, Adam and Brown, Tom B and Song, Dawn and Erlingsson, Ulfar and Oprea, Alina and Raffel, Colin},
	year = {2021},
}

@misc{carlini_quantifying_2022,
	title = {Quantifying {Memorization} {Across} {Neural} {Language} {Models}},
	doi = {10.48550/arXiv.2202.07646},
	author = {Carlini, Nicholas and Ippolito, Daphne and Jagielski, Matthew and Lee, Katherine and Tramer, Florian and Zhang, Chiyuan},
	month = feb,
	year = {2022},
}

@inproceedings{winkler_string_1990,
	title = {String {Comparator} {Metrics} and {Enhanced} {Decision} {Rules} in the {Fellegi}-{Sunter} {Model} of {Record} {Linkage}},
	booktitle = {Proceedings of the {Section} on {Survey} {Research} {Methods}},
	author = {Winkler, William E.},
	year = {1990},
	pages = {354--359},
}

@article{levenshtein_binary_1966,
	title = {Binary codes capable of correcting deletions, insertions and reversals},
	volume = {10},
	journal = {Soviet Physics Doklady},
	author = {Levenshtein, Vladimir Iosifovich},
	month = feb,
	year = {1966},
	pages = {707--710},
}

@article{hamming_error_1950,
	title = {Error detecting and error correcting codes},
	volume = {29},
	doi = {10.1002/j.1538-7305.1950.tb00463.x},
	number = {2},
	journal = {The Bell System Technical Journal},
	author = {Hamming, R. W.},
	year = {1950},
	pages = {147--160},
}

@article{wang_variational_2022,
	title = {Variational {Model} {Inversion} {Attacks}},
	volume = {34},
	doi = {10.48550/arXiv.2201.10787},
	journal = {NeurIPS},
	author = {Wang, Kuan-Chieh and Fu, Yan and Li, Ke and Khisti, Ashish and Zemel, Richard and Makhzani, Alireza},
	month = jan,
	year = {2022},
}

@inproceedings{vaswani_attention_2017,
	title = {Attention {Is} {All} {You} {Need}},
	doi = {10.48550/arXiv.1706.03762},
	booktitle = {{NeurIPS}},
	author = {Vaswani, Ashish and Shazeer, Noam and Parmar, Niki and Uszkoreit, Jakob and Jones, Llion and Gomez, Aidan N. and Kaiser, Lukasz and Polosukhin, Illia},
	year = {2017},
}

@inproceedings{luo_geolayoutlm_2023,
	title = {{GeoLayoutLM}: {Geometric} {Pre}-training for {Visual} {Information} {Extraction}},
	doi = {10.48550/arXiv.2304.10759},
	booktitle = {{IEEE}/{CVF} {CVPR}},
	author = {Luo, Chuwei and Cheng, Changxu and Zheng, Qi and Yao, Cong},
	month = jun,
	year = {2023},
}

@inproceedings{reimers_sentence-bert_2019,
	title = {Sentence-{BERT}: {Sentence} {Embeddings} using {Siamese} {BERT}-{Networks}},
	doi = {10.18653/v1/D19-1410},
	booktitle = {{ACL}-{EMNLP}-{IJCNLP}},
	author = {Reimers, Nils and Gurevych, Iryna},
	year = {2019},
}

@inproceedings{perez_red_2022,
	title = {Red {Teaming} {Language} {Models} with {Language} {Models}},
	doi = {10.18653/v1/2022.emnlp-main.225},
	booktitle = {{EMNLP}},
	author = {Perez, Ethan and Huang, Saffron and Song, Francis and Cai, Trevor and Ring, Roman and Aslanides, John and Glaese, Amelia and McAleese, Nat and Irving, Geoffrey},
	year = {2022},
}

@inproceedings{parikh_canary_2022,
	title = {Canary {Extraction} in {Natural} {Language} {Understanding} {Models}},
	volume = {2},
	doi = {10.18653/v1/2022.acl-short.61},
	booktitle = {Annual {Meeting} of the {ACL}},
	author = {Parikh, Rahil and Dupuy, Christophe and Gupta, Rahul},
	year = {2022},
	pages = {552--560},
}

@inproceedings{mireshghallah_quantifying_2022,
	title = {Quantifying {Privacy} {Risks} of {Masked} {Language} {Models} {Using} {Membership} {Inference} {Attacks}},
	doi = {10.18653/v1/2022.emnlp-main.570},
	booktitle = {{ACL}-{EMNLP}},
	author = {Mireshghallah, Fatemehsadat and Goyal, Kartik and Uniyal, Archit and Berg-Kirkpatrick, Taylor and Shokri, Reza},
	month = nov,
	year = {2022},
}

@inproceedings{shokri_membership_2017,
	title = {Membership {Inference} {Attacks} against {Machine} {Learning} {Models}},
	doi = {10.1109/SP.2017.41},
	booktitle = {{IEEE} {S}\&{P}},
	author = {Shokri, Reza and Stronati, Marco and Song, Congzheng and Shmatikov, Vitaly},
	year = {2017},
}

@inproceedings{salem_let_2023,
	title = {Let the {Privacy} {Games} {Begin}! {A} {Unified} {Treatment} of {Data} {Inference} {Privacy} in {Machine} {Learning}},
	shorttitle = {{SoK}},
	doi = {10.1109/SP46215.2023.10179281},
	booktitle = {{IEEE} {S}\&{P}},
	author = {Salem, Ahmed and Cherubin, Giovanni and Evans, David and Köpf, Boris and Paverd, Andrew and Suri, Anshuman and Tople, Shruti and Zanella-Béguelin, Santiago},
	month = may,
	year = {2023},
}

@inproceedings{rezaei_accuracy-privacy_2023,
	title = {Accuracy-{Privacy} {Trade}-off in {Deep} {Ensemble}: {A} {Membership} {Inference} {Perspective}},
	shorttitle = {Accuracy-{Privacy} {Trade}-off in {Deep} {Ensemble}},
	doi = {10.1109/SP46215.2023.10179463},
	booktitle = {{IEEE} {S}\&{P}},
	author = {Rezaei, Shahbaz and Shafiq, Zubair and Liu, Xin},
	month = may,
	year = {2023},
	pages = {364--381},
}

@inproceedings{lukas_analyzing_2023,
	title = {Analyzing {Leakage} of {Personally} {Identifiable} {Information} in {Language} {Models}},
	doi = {10.1109/SP46215.2023.10179300},
	booktitle = {{IEEE} {S}\&{P}},
	author = {Lukas, Nils and Salem, Ahmed and Sim, Robert and Tople, Shruti and Wutschitz, Lukas and Zanella-Béguelin, Santiago},
	month = may,
	year = {2023},
}

@inproceedings{zhang_secret_2020,
	title = {The {Secret} {Revealer}: {Generative} {Model}-{Inversion} {Attacks} {Against} {Deep} {Neural} {Networks}},
	shorttitle = {The {Secret} {Revealer}},
	doi = {10.1109/CVPR42600.2020.00033},
	booktitle = {{IEEE}/{CVF} {CVPR}},
	author = {Zhang, Yuheng and Jia, Ruoxi and Pei, Hengzhi and Wang, Wenxiao and Li, Bo and Song, Dawn},
	year = {2020},
}

@inproceedings{kahla_label-only_2022,
	title = {Label-{Only} {Model} {Inversion} {Attacks} via {Boundary} {Repulsion}},
	doi = {10.1109/CVPR52688.2022.01462},
	booktitle = {{IEEE}/{CVF} {CVPR}},
	author = {Kahla, Mostafa and Chen, Si and Just, Hoang Anh and Jia, Ruoxi},
	month = jun,
	year = {2022},
	pages = {15025--15033},
}

@inproceedings{yu_bag_2023,
	title = {Bag of {Tricks} for {Training} {Data} {Extraction} from {Language} {Models}},
	doi = {10.48550/arXiv.2302.04460},
	booktitle = {{ICML}},
	author = {Yu, Weichen and Pang, Tianyu and Liu, Qian and Du, Chao and Kang, Bingyi and Huang, Yan and Lin, Min and Yan, Shuicheng},
	month = jun,
	year = {2023},
}

@inproceedings{kandpal_deduplicating_2022,
	title = {Deduplicating {Training} {Data} {Mitigates} {Privacy} {Risks} in {Language} {Models}},
	volume = {162},
	doi = {10.48550/arXiv.2202.06539},
	booktitle = {{ICML}},
	author = {Kandpal, Nikhil and Wallace, Eric and Raffel, Colin},
	month = jul,
	year = {2022},
	pages = {10697--10707},
}

@inproceedings{holtzman_curious_2019,
	title = {The {Curious} {Case} of {Neural} {Text} {Degeneration}},
	doi = {10.48550/arXiv.1904.09751},
	doi = {10.48550/arXiv.1904.09751},
	booktitle = {{ICLR}},
	author = {Holtzman, Ari and Buys, Jan and Du, Li and Forbes, Maxwell and Choi, Yejin},
	month = apr,
	year = {2019},
}

@inproceedings{goyal_exposing_2022,
	title = {Exposing the {Implicit} {Energy} {Networks} behind {Masked} {Language} {Models} via {Metropolis}--{Hastings}},
	doi = {10.48550/arXiv.2106.02736},
	booktitle = {{ICLR}},
	author = {Goyal, Kartik and Dyer, Chris and Berg-Kirkpatrick, Taylor},
	month = mar,
	year = {2022},
}

@inproceedings{he_mask_2017,
	title = {Mask {R}-{CNN}},
	doi = {10.1109/ICCV.2017.322},
	booktitle = {{IEEE} {ICCV}},
	author = {He, Kaiming and Gkioxari, Georgia and Dollar, Piotr and Girshick, Ross},
	month = oct,
	year = {2017},
	pages = {2980--2988},
}

@inproceedings{appalaraju_docformer_2021,
	title = {{DocFormer}: {End}-to-{End} {Transformer} for {Document} {Understanding}},
	shorttitle = {{DocFormer}},
	doi = {10.48550/arXiv.2106.11539},
	booktitle = {{IEEE}/{CVF} {ICCV}},
	author = {Appalaraju, Srikar and Jasani, Bhavan and Kota, Bhargava Urala and Xie, Yusheng and Manmatha, R.},
	month = oct,
	year = {2021},
}

@misc{elmahdy_deconstructing_2023,
	title = {Deconstructing {Classifiers}: {Towards} {A} {Data} {Reconstruction} {Attack} {Against} {Text} {Classification} {Models}},
	doi = {10.48550/arXiv.2306.13789},
	author = {Elmahdy, Adel and Salem, Ahmed},
	month = jun,
	year = {2023},
}

@inproceedings{wang_bert_2019,
	title = {{BERT} has a {Mouth}, and {It} {Must} {Speak}: {BERT} as a {Markov} {Random} {Field} {Language} {Model}},
	doi = {10.18653/v1/W19-2304},
	booktitle = {{ACL} {W19}-23},
	author = {Wang, Alex and Cho, Kyunghyun},
	month = jun,
	year = {2019},
	pages = {30--36},
}

@inproceedings{song_information_2020,
	title = {Information {Leakage} in {Embedding} {Models}},
	doi = {10.1145/3372297.3417270},
	booktitle = {{ACM} {SIGSAC} {CCS}},
	author = {Song, Congzheng and Raghunathan, Ananth},
	month = oct,
	year = {2020},
	pages = {377--390},
}

@inproceedings{mireshghallah_privacy_2021,
	title = {Privacy {Regularization}: {Joint} {Privacy}-{Utility} {Optimization} in {LanguageModels}},
	doi = {10.18653/v1/2021.naacl-main.298},
	booktitle = {{NAACL}},
	author = {Mireshghallah, Fatemehsadat and Inan, Huseyin and Hasegawa, Marcello and Rühle, Victor and Berg-Kirkpatrick, Taylor and Sim, Robert},
	month = jun,
	year = {2021},
}

@inproceedings{meehan_sentence-level_2022,
	title = {Sentence-level {Privacy} for {Document} {Embeddings}},
	doi = {10.18653/v1/2022.acl-long.238},
	booktitle = {Annual {Meeting} of the {ACL}},
	author = {Meehan, Casey and Mrini, Khalil and Chaudhuri, Kamalika},
	year = {2022},
}

@inproceedings{mathew_docvqa_2021,
	title = {{DocVQA}: {A} {Dataset} for {VQA} on {Document} {Images}},
	doi = {10.48550/arXiv.2007.00398},
	booktitle = {{IEEE}/{CVF} {WACV}},
	author = {Mathew, Minesh and Karatzas, Dimosthenis and Jawahar, C. V.},
	month = jan,
	year = {2021},
}

@inproceedings{lehman_does_2021,
	title = {Does {BERT} {Pretrained} on {Clinical} {Notes} {Reveal} {Sensitive} {Data}?},
	doi = {10.18653/v1/2021.naacl-main.73},
	booktitle = {{NAACL}},
	author = {Lehman, Eric and Jain, Sarthak and Pichotta, Karl and Goldberg, Yoav and Wallace, Byron},
	year = {2021},
}

@inproceedings{lee_language_2022,
	title = {Do {Language} {Models} {Plagiarize}?},
	doi = {10.1145/3543507.3583199},
	booktitle = {{ACM} {WWW}},
	author = {Lee, Jooyoung and Le, Thai and Chen, Jinghui and Lee, Dongwon},
	month = mar,
	year = {2022},
	pages = {3637--3647},
}

@inproceedings{hong_bros_2021,
	title = {{BROS}: {A} {Pre}-trained {Language} {Model} {Focusing} on {Text} and {Layout} for {Better} {Key} {Information} {Extraction} from {Documents}},
	doi = {10.1609/aaai.v36i10.21322},
	booktitle = {{AAAI}},
	author = {Hong, Teakgyu and Kim, Donghyun and Ji, Mingi and Hwang, Wonseok and Nam, Daehyun and Park, Sungrae},
	month = aug,
	year = {2021},
}

@inproceedings{ficler_controlling_2017,
	title = {Controlling {Linguistic} {Style} {Aspects} in {Neural} {Language} {Generation}},
	doi = {10.18653/v1/W17-4912},
	booktitle = {Style-{Var}},
	author = {Ficler, Jessica and Goldberg, Yoav},
	year = {2017},
	pages = {94--104},
}

@inproceedings{elmahdy_privacy_2022,
	title = {Privacy {Leakage} in {Text} {Classification} {A} {Data} {Extraction} {Approach}},
	doi = {10.18653/v1/2022.privatenlp-1.3},
	booktitle = {{PrivateNLP}},
	author = {Elmahdy, Adel and A. Inan, Huseyin and Sim, Robert},
	year = {2022},
}

@inproceedings{continella_obfuscation-resilient_2017,
	title = {Obfuscation-{Resilient} {Privacy} {Leak} {Detection} for {Mobile} {Apps} {Through} {Differential} {Analysis}},
	doi = {10.14722/ndss.2017.23465},
	booktitle = {{NDSS}},
	author = {Continella, Andrea and Fratantonio, Yanick and Lindorfer, Martina and Puccetti, Alessandro and Zand, Ali and Kruegel, Christopher and Vigna, Giovanni},
	year = {2017},
}

@inproceedings{budzianowski_hello_2019,
	title = {Hello, {It}’s {GPT}-2 - {How} {Can} {I} {Help} {You}? {Towards} the {Use} of {Pretrained} {Language} {Models} for {Task}-{Oriented} {Dialogue} {Systems}},
	shorttitle = {Hello, {It}’s {GPT}-2 - {How} {Can} {I} {Help} {You}?},
	doi = {10.18653/v1/D19-5602},
	booktitle = {{NGT}},
	author = {Budzianowski, Paweł and Vulić, Ivan},
	year = {2019},
	pages = {15--22},
}

@inproceedings{llados_lambert_2021,
	title = {{LAMBERT}: {Layout}-{Aware} {Language} {Modeling} for {Information} {Extraction}},
	shorttitle = {{LAMBERT}},
	doi = {10.48550/arXiv.2002.08087},
	booktitle = {{ICDAR}},
	author = {Garncarek, Łukasz and Powalski, Rafał and Stanisławek, Tomasz and Topolski, Bartosz and Halama, Piotr and Turski, Michał and Graliński, Filip},
	year = {2021},
}

@inproceedings{devlin_bert_2019,
	title = {{BERT}: {Pre}-training of {Deep} {Bidirectional} {Transformers} for {Language} {Understanding}},
	doi = {10.18653/v1/N19-1423},
	booktitle = {{NAACL}},
	author = {Devlin, Jacob and Chang, Ming-Wei and Lee, Kenton and Toutanova, Kristina},
	month = jun,
	year = {2019},
}

@article{zhang_understanding_2021,
	title = {Understanding deep learning (still) requires rethinking generalization},
	volume = {64},
	doi = {10.1145/3446776},
	journal = {CACM},
	author = {Zhang, Chiyuan and Bengio, Samy and Hardt, Moritz and Recht, Benjamin and Vinyals, Oriol},
	month = mar,
	year = {2021},
}

@inproceedings{yeom_privacy_2018,
	title = {Privacy {Risk} in {Machine} {Learning}: {Analyzing} the {Connection} to {Overfitting}},
	doi = {10.1109/CSF.2018.00027},
	booktitle = {{IEEE} {CSF}},
	author = {Yeom, Samuel and Giacomelli, Irene and Fredrikson, Matt and Jha, Somesh},
	year = {2018},
	pages = {268--282},
}

@inproceedings{xu_layoutlm_2019,
	title = {{LayoutLM}: {Pre}-training of {Text} and {Layout} for {Document} {Image} {Understanding}},
	doi = {10.1145/3394486.3403172},
	booktitle = {{ACM} {SIGKDD}},
	author = {Xu, Yiheng and Li, Minghao and Cui, Lei and Huang, Shaohan and Wei, Furu and Zhou, Ming},
	month = dec,
	year = {2019},
}

@inproceedings{xu_layoutlmv2_2021,
	title = {{LayoutLMv2}: {Multi}-modal {Pre}-training for {Visually}-rich {Document} {Understanding}},
	doi = {10.18653/v1/2021.acl-long.201},
	booktitle = {{ACL}-{IJCNLP}},
	author = {Xu, Yang and Xu, Yiheng and Lv, Tengchao and Cui, Lei and Wei, Furu and Wang, Guoxin and Lu, Yijuan and Florencio, Dinei and Zhang, Cha and Che, Wanxiang and Zhang, Min and Zhou, Lidong},
	year = {2021},
}

@inproceedings{wu_methodology_2016,
	title = {A {Methodology} for {Formalizing} {Model}-{Inversion} {Attacks}},
	doi = {10.1109/CSF.2016.32},
	booktitle = {{IEEE} {CSF}},
	author = {Wu, Xi and Fredrikson, Matthew and Jha, Somesh and Naughton, Jeffrey F.},
	month = jun,
	year = {2016},
	pages = {355--370},
}

@inproceedings{vakili_downstream_2022,
	title = {Downstream {Task} {Performance} of {BERT} {Models} {Pre}-{Trained} {Using} {Automatically} {De}-{Identified} {Clinical} {Data}},
	url = {https://aclanthology.org/2022.lrec-1.451},
	booktitle = {{LREC}},
	author = {Vakili, Thomas and Lamproudis, Anastasios and Henriksson, Aron and Dalianis, Hercules},
	month = jun,
	year = {2022},
}

@inproceedings{ren_recon_2016,
	title = {{ReCon}: {Revealing} and {Controlling} {PII} {Leaks} in {Mobile} {Network} {Traffic}},
	shorttitle = {{ReCon}},
	doi = {10.1145/2906388.2906392},
	booktitle = {{ACM} {MobiSys}},
	author = {Ren, Jingjing and Rao, Ashwin and Lindorfer, Martina and Legout, Arnaud and Choffnes, David},
	month = jun,
	year = {2016},
}

@article{raffel_exploring_2020,
	title = {Exploring the {Limits} of {Transfer} {Learning} with a {Unified} {Text}-to-{Text} {Transformer}},
	url = {http://jmlr.org/papers/v21/20-074.html},
	journal = {JMLR},
	author = {Raffel, Colin and Shazeer, Noam and Roberts, Adam and Lee, Katherine and Narang, Sharan and Matena, Michael and Zhou, Yanqi and Li, Wei and Liu, Peter J.},
	year = {2020},
}

@inproceedings{powalski_going_2021,
	title = {Going {Full}-{TILT} {Boogie} on {Document} {Understanding} with {Text}-{Image}-{Layout} {Transformer}},
	doi = {10.1007/978-3-030-86331-9_47},
	booktitle = {{ICDAR}},
	author = {Powalski, Rafał and Borchmann, Łukasz and Jurkiewicz, Dawid and Dwojak, Tomasz and Pietruszka, Michał and Pałka, Gabriela},
	month = jul,
	year = {2021},
}

@inproceedings{olatunji_membership_2021,
	title = {Membership {Inference} {Attack} on {Graph} {Neural} {Networks}},
	doi = {10.1109/TPSISA52974.2021.00002},
	booktitle = {{IEEE} {TPS}-{ISA}},
	author = {Olatunji, Iyiola E. and Nejdl, Wolfgang and Khosla, Megha},
	month = dec,
	year = {2021},
}

@article{nikolentzos_message_2020,
	title = {Message {Passing} {Attention} {Networks} for {Document} {Understanding}},
	doi = {10.1609/aaai.v34i05.6376},
	journal = {AAAI},
	author = {Nikolentzos, Giannis and Tixier, Antoine and Vazirgiannis, Michalis},
	month = apr,
	year = {2020},
}

@inproceedings{lewis_building_2006,
	title = {Building a test collection for complex document information processing},
	doi = {10.1145/1148170.1148307},
	booktitle = {{ACM} {SIGIR}},
	author = {Lewis, D. and Agam, G. and Argamon, S. and Frieder, O. and Grossman, D. and Heard, J.},
	month = aug,
	year = {2006},
}

@article{khosravy_model_2022,
	title = {Model {Inversion} {Attack} by {Integration} of {Deep} {Generative} {Models}: {Privacy}-{Sensitive} {Face} {Generation} {From} a {Face} {Recognition} {System}},
	doi = {10.1109/TIFS.2022.3140687},
	journal = {IEEE TIFS},
	author = {Khosravy, Mahdi and Nakamura, Kazuaki and Hirose, Yuki and Nitta, Naoko and Babaguchi, Noboru},
	year = {2022},
}

@inproceedings{jaume_funsd_2019,
	title = {{FUNSD}: {A} {Dataset} for {Form} {Understanding} in {Noisy} {Scanned} {Documents}},
	doi = {10.1109/ICDARW.2019.10029},
	booktitle = {{ICDAR}},
	author = {Jaume, Guillaume and Ekenel, Hazim Kemal and Thiran, Jean-Philippe},
	month = may,
	year = {2019},
}

@article{jaro_advances_1989,
	title = {Advances in {Record}-{Linkage} {Methodology} as {Applied} to {Matching} the 1985 {Census} of {Tampa}, {Florida}},
	doi = {10.1080/01621459.1989.10478785},
	journal = {Journal of the ASA},
	author = {Jaro, Matthew A.},
	month = jun,
	year = {1989},
}

@inproceedings{hwang_spatial_2021,
	title = {Spatial {Dependency} {Parsing} for {Semi}-{Structured} {Document} {Information} {Extraction}},
	doi = {10.18653/v1/2021.findings-acl.28},
	booktitle = {{ACL}-{IJCNLP}},
	author = {Hwang, Wonseok and Yim, Jinyeong and Park, Seunghyun and Yang, Sohee and Seo, Minjoon},
	year = {2021},
}

@inproceedings{huang_icdar2019_2019,
	title = {{ICDAR2019} {Competition} on {Scanned} {Receipt} {OCR} and {Information} {Extraction}},
	doi = {10.1109/ICDAR.2019.00244},
	booktitle = {{ICDAR}},
	author = {Huang, Zheng and Chen, Kai and He, Jianhua and Bai, Xiang and Karatzas, Dimosthenis and Lu, Shijian and Jawahar, C. V.},
	month = sep,
	year = {2019},
	pages = {1516--1520},
}

@inproceedings{huang_layoutlmv3_2022,
	title = {{LayoutLMv3}: {Pre}-training for {Document} {AI} with {Unified} {Text} and {Image} {Masking}},
	doi = {10.1145/3503161.3548112},
	booktitle = {{ACM} {Multimedia}},
	author = {Huang, Yupan and Lv, Tengchao and Cui, Lei and Lu, Yutong and Wei, Furu},
	month = apr,
	year = {2022},
	pages = {4083--4091},
}

@article{hu_membership_2022,
	title = {Membership {Inference} {Attacks} on {Machine} {Learning}: {A} {Survey}},
	shorttitle = {Membership {Inference} {Attacks} on {Machine} {Learning}},
	doi = {10.1145/3523273},
	journal = {ACM Surveys},
	author = {Hu, Hongsheng and Salcic, Zoran and Sun, Lichao and Dobbie, Gillian and Yu, Philip S. and Zhang, Xuyun},
	month = jan,
	year = {2022},
}

@inproceedings{harley_evaluation_2015,
	title = {Evaluation of {Deep} {Convolutional} {Nets} for {Document} {Image} {Classification} and {Retrieval}},
	doi = {10.1109/ICDAR.2015.7333910},
	booktitle = {{ICDAR}},
	author = {Harley, Adam W. and Ufkes, Alex and Derpanis, Konstantinos G.},
	month = feb,
	year = {2015},
	pages = {991--995},
}

@inproceedings{guo_eaten_2019,
	title = {{EATEN}: {Entity}-{Aware} {Attention} for {Single} {Shot} {Visual} {Text} {Extraction}},
	shorttitle = {{EATEN}},
	doi = {10.1109/ICDAR.2019.00049},
	booktitle = {{ICDAR}},
	author = {Guo, He and Qin, Xiameng and Liu, Jiaming and Han, Junyu and Liu, Jingtuo and Ding, Errui},
	month = sep,
	year = {2019},
}

@inproceedings{dhouib_docparser_2023,
	title = {{DocParser}: {End}-to-end {OCR}-free {Information} {Extraction} from {Visually} {Rich} {Documents}},
	shorttitle = {{DocParser}},
	doi = {10.48550/arXiv.2304.12484},
	booktitle = {{ICDAR}},
	author = {Dhouib, Mohamed and Bettaieb, Ghassen and Shabou, Aymen},
	month = may,
	year = {2023},
}

@inproceedings{fredrikson_privacy_2014,
	title = {Privacy in {Pharmacogenetics}: {An} {End}-to-{End} {Case} {Study} of {Personalized} {Warfarin} {Dosing}},
	booktitle = {{USENIX} {Security}},
	author = {Fredrikson, Matthew and Lantz, Eric and Jha, Somesh and Lin, Simon and Page, David and Ristenpart, Thomas},
	year = {2014},
}

@inproceedings{fredrikson_model_2015,
	title = {Model {Inversion} {Attacks} that {Exploit} {Confidence} {Information} and {Basic} {Countermeasures}},
	doi = {10.1145/2810103.2813677},
	booktitle = {{ACM} {SIGSAC} {CCS}},
	author = {Fredrikson, Matt and Jha, Somesh and Ristenpart, Thomas},
	year = {2015},
	pages = {1322--1333},
}

@inproceedings{feldman_does_2020,
	title = {Does learning require memorization? a short tale about a long tail},
	doi = {10.1145/3357713.3384290},
	booktitle = {{ACM} {SIGACT} {STOC}},
	author = {Feldman, Vitaly},
	month = jun,
	year = {2020},
	pages = {954--959},
}

\section*{Appendix}
\renewcommand{\thesubsection}{\Alph{subsection}}

\subsection{Training conditions}

This section elaborates on aspects discussed in section \ref{sec:dataset_training_conditions}.

First, table \ref{tab:volume_datasets} shows the volumes of the datasets we used. We used two datasets: FUNSD \cite{jaume_funsd_2019} and SROIE \cite{huang_icdar2019_2019}, which we split into three non-overlapping parts: validation, train-public, and train-private. Column 3 displays the number of documents in each part. Columns 4 and 5 present the total number of fields in each dataset, and the number of fields we selected for our attack, respectively.

Second, the results of our hyperparameter optimization are as follows. For step 1, $N_c = 128$ offers a satisfactory balance between performance and computational time ($N_c = 512$ marginally improves the attack accuracy by 2\%, at a significant computational cost). For steps 3 and 4, effective results were typically associated with a low temperature, about 0.2--0.4, with a decay rate applied over 3 steps for the PUB-MLM softmax, and no decay for the target one. For step 5, geometric mean exhibited superior results, though the optimal weight varied considerably across configurations (between 0.2 and 0.6). For step 6, the optimal parameter was $p=0.10$, except for some settings where $p=0.12$ proved superior. Finally, we selected $n_a = 8$ attempts for the multi-shot variant.

\begin{table}[t!]
\begin{center}
\begin{tabular}{ccccc}
\hline
Dataset & Partition      & \# doc & \# field & \# selected \\ \hline
FUNSD   & Valid     & 50     & 809      & 495          \\
FUNSD   & Train-PUB & 74     & 1436     & 911          \\
FUNSD   & Train-PRI & 75     & 1296     & 777          \\
SROIE   & Valid     & 100    & 400      & 301          \\
SROIE   & Train-PUB & 263    & 1052     & 796          \\
SROIE   & Train-PRI & 263    & 1052     & 791          \\ \hline
\end{tabular}
\caption{Number of documents and fields in our partitions.}
\label{tab:volume_datasets}
\end{center}
\end{table}

\begin{table*}[t!]
\centering
\begin{tabular}{|@{\hspace{5pt}}c@{\hspace{5pt}}|@{\hspace{5pt}}c@{\hspace{5pt}}c@{\hspace{5pt}}c@{\hspace{5pt}}c@{\hspace{5pt}}|@{\hspace{5pt}}c@{\hspace{5pt}}|@{\hspace{5pt}}c@{\hspace{5pt}}|@{\hspace{5pt}}c@{\hspace{5pt}}c@{\hspace{5pt}}c@{\hspace{5pt}}c@{\hspace{5pt}}c@{\hspace{5pt}}|}
\hline
\textbf{Backbone} & \textbf{Task} & \textbf{Data} & \textbf{Crit} & \textbf{Modal} & \textbf{Prec} & \textbf{IpF} & \textbf{HamAAC} & \textbf{AccAUC} & \textbf{AccAt 1\%} & \textbf{AccAt 5\%} & \textbf{AccAt 100\%} \\ \hline
LayoutLM & MLM & FUN & Prec & UniM & 0.545 & 1.132 & 0.153 & 0.020 & 0.125 & 0.051 & 0.009 \\
LayoutLM & MLM & FUN & Prec & BiM & 0.545 & \underline{1.395} & \underline{0.274} & 0.061 & 0.125 & 0.154 & 0.027 \\
LayoutLM & MLM & FUN & Loss & UniM & 0.543 & 1.123 & 0.198 & 0.021 & 0.125 & 0.077 & 0.010 \\
LayoutLM & MLM & FUN & Loss & BiM & 0.543 & 1.171 & 0.181 & 0.034 & \underline{\textbf{0.250}} & 0.103 & 0.013 \\
LayoutLM & MLM & SRO & Prec & UniM & 0.540 & 1.081 & 0.104 & 0.036 & 0.000 & 0.075 & 0.020 \\
LayoutLM & MLM & SRO & Prec & BiM & 0.538 & 1.393 & 0.191 & \underline{0.070} & 0.125 & \underline{\textbf{0.225}} & \underline{0.032} \\
LayoutLM & MLM & SRO & Loss & UniM & 0.534 & 0.977 & 0.077 & 0.017 & 0.000 & 0.025 & 0.010 \\
LayoutLM & MLM & SRO & Loss & BiM & 0.534 & 1.226 & 0.112 & 0.031 & 0.000 & 0.025 & 0.021 \\ \hline
LayoutLM & EE-BIO & FUN & Prec & UniM & 0.762 & 1.075 & 0.216 & 0.016 & \underline{0.125} & \underline{0.026} & 0.008 \\
LayoutLM & EE-BIO & FUN & Prec & BiM & 0.746 & 1.024 & 0.201 & 0.004 & 0.000 & 0.000 & 0.004 \\
LayoutLM & EE-BIO & FUN & Loss & UniM & 0.718 & 1.053 & 0.203 & 0.008 & \underline{0.125} & \underline{0.026} & 0.004 \\
LayoutLM & EE-BIO & FUN & Loss & BiM & 0.702 & 1.024 & 0.188 & 0.003 & 0.000 & 0.000 & 0.001 \\
LayoutLM & EE-BIO & SRO & Prec & UniM & 0.960 & 1.252 & 0.231 & \underline{0.026} & 0.000 & 0.025 & \underline{0.019} \\
LayoutLM & EE-BIO & SRO & Prec & BiM & 0.967 & 1.207 & 0.204 & 0.010 & 0.000 & 0.000 & 0.010 \\
LayoutLM & EE-BIO & SRO & Loss & UniM & 0.940 & 1.246 & \underline{0.249} & 0.023 & 0.000 & 0.025 & 0.018 \\
LayoutLM & EE-BIO & SRO & Loss & BiM & 0.955 & \underline{1.294} & 0.231 & 0.014 & 0.000 & 0.000 & 0.013 \\ \hline
LayoutLM & EE-SPD & FUN & Prec & UniM & 0.760 & 1.054 & 0.177 & 0.005 & 0.000 & 0.000 & 0.004 \\
LayoutLM & EE-SPD & FUN & Prec & BiM & 0.760 & 1.001 & 0.181 & 0.008 & 0.000 & 0.026 & 0.004 \\
LayoutLM & EE-SPD & FUN & Loss & UniM & 0.469 & 1.036 & 0.144 & 0.012 & \underline{0.125} & 0.026 & 0.004 \\
LayoutLM & EE-SPD & FUN & Loss & BiM & 0.623 & 0.979 & 0.170 & 0.007 & \underline{0.125} & 0.026 & 0.001 \\
LayoutLM & EE-SPD & SRO & Prec & UniM & 0.952 & 1.141 & 0.213 & 0.019 & 0.000 & 0.050 & 0.010 \\
LayoutLM & EE-SPD & SRO & Prec & BiM & 0.935 & 1.209 & 0.192 & 0.008 & 0.000 & 0.000 & 0.009 \\
LayoutLM & EE-SPD & SRO & Loss & UniM & 0.856 & \underline{1.224} & \underline{0.227} & \underline{0.044} & \underline{0.125} & \underline{0.150} & \underline{0.020} \\
LayoutLM & EE-SPD & SRO & Loss & BiM & 0.880 & 1.221 & 0.205 & 0.009 & 0.000 & 0.050 & 0.005 \\ \hline
LayoutLM & EL & FUN & Prec & UniM & 0.240 & \underline{1.164} & \underline{0.135} & \underline{0.012} & 0.000 & \underline{0.026} & \underline{0.013} \\
LayoutLM & EL & FUN & Prec & BiM & 0.193 & 0.986 & 0.059 & 0.005 & 0.000 & \underline{0.026} & 0.001 \\
LayoutLM & EL & FUN & Loss & UniM & 0.093 & 1.019 & 0.118 & 0.006 & \underline{0.125} & \underline{0.026} & 0.001 \\
LayoutLM & EL & FUN & Loss & BiM & 0.000 & 1.015 & 0.116 & 0.002 & 0.000 & 0.000 & 0.001 \\ \hline
BROS & MLM & FUN & Prec & UniM & 0.546 & 1.243 & 0.302 & 0.051 & \underline{\textbf{0.250}} & 0.128 & 0.021 \\
BROS & MLM & FUN & Prec & BiM & 0.546 & \underline{\textbf{1.540}} & \underline{\textbf{0.389}} & \underline{\textbf{0.078}} & \underline{\textbf{0.250}} & \underline{0.154} & \underline{\textbf{0.041}} \\
BROS & MLM & FUN & Loss & UniM & 0.543 & 1.170 & 0.276 & 0.038 & \underline{0.125} & 0.077 & 0.019 \\
BROS & MLM & FUN & Loss & BiM & 0.539 & 1.246 & 0.312 & 0.041 & \underline{0.125} & 0.051 & 0.023 \\
BROS & MLM & SRO & Prec & UniM & 0.541 & 1.028 & 0.102 & 0.007 & \underline{0.125} & 0.025 & 0.001 \\
BROS & MLM & SRO & Prec & BiM & 0.539 & 1.092 & 0.126 & 0.008 & 0.000 & 0.000 & 0.006 \\
BROS & MLM & SRO & Loss & UniM & 0.534 & 1.017 & 0.072 & 0.000 & 0.000 & 0.000 & 0.000 \\
BROS & MLM & SRO & Loss & BiM & 0.534 & 1.001 & 0.080 & 0.000 & 0.000 & 0.000 & 0.000 \\ \hline
\end{tabular}
\caption{Main experimental results. Columns 1-5 show the target's model training conditions: backbone (LayoutLM, BROS), task (MLM, EE-BIO, EE-SPADE, EL), dataset (FUNSD, SROIE), criterion (Precision, Loss), modality (Unimodal, Bimodal). Column 6 shows its validation accuracy. Columns 7 and 8-12 show the results of the one-shot and multi-shot attacks, respectively.}
\label{tab:detailed_results}
\end{table*}

\subsection{Evaluation of the multi-shot variant}

Let us detail further $AccAUC$ and $HamAAC$ metrics introduced in section \ref{sec:eval_multishot}. For $p \in [0, 1]$, let $T(p)$ be the set of field that are ranked top-$p$ by the membership inference metric, and let $Ham$ be the Hamming distance \cite{hamming_error_1950}. We define:

\begin{gather}
    AccAt(p) := \frac{1}{|T(p)|} \sum_{f \in T(p)} \mathbb{1}(f = \widetilde{f}) \ \ \in [0, 1] \\
    AccAUC := \int_{p=0}^{1} AccAt(p) \diff p \ \  \in [0, 1] \\ 
    HamAt(p) := \frac{1}{|T(p)|} \sum_{f \in T(p)} Ham(f, \widetilde{f}) \ \ \in [0, 1] \\
    HamAAC := 1 - \int_{p=0}^{1} HamAt(p) \diff p \ \ \in [0, 1]
\end{gather}

Let us now prove that $HamAAC$ satisfies the two properties of section \ref{sec:eval_multishot}; the demonstration is similar for $AccAUC$:

\begin{enumerate}
    \item Let us fix the values of all $f$ and $\widetilde{f}$. Then, let be two indices $1 \le i < j \le M$ such that $Ham(f_i, \widetilde{f_i}) > Ham(f_j, \widetilde{f_j})$. If we swap $i$ and $j$, this will increase $HamAt(p)$ for $p \in [i/M, j/M[$ leaving the other values unchanged, which will increase the $HamAAC$ score. As a result, every step of a selection sort applied to sort the indices by increasing values of $Ham(f_i, \widetilde{f_i})$ will increase $HamAAC$, proving that the maximum will be achieved at the end of the selection sort.
    \item Let $F$ and $\widetilde{F}$ have fixed order, and length $M$, and let $i \in \llbracket 1, M \rrbracket$. Then, if we change the reconstruction $\widetilde{f_i}$ into another $\widetilde{f_i'}$ such that $Ham(f_i, \widetilde{f_i'}) > Ham(f_i, \widetilde{f_i})$, it is clear that $\forall p \ge i / M$, $HamAt(p)$ increases. As a result, this improvement of $\widetilde{f_i}$ increases $HamAAC$ as well.
 \end{enumerate}

\subsection{Detailed experimental results}

Table \ref{tab:detailed_results} details our experimental results, using $N_c = 128$ candidates. We excluded non-MLM tasks with BROS backbone because of their lower results (see section \ref{sec:experimental_perfs}).

\end{document}